\documentclass[aps, notitlepage,nofootinbib,superscriptaddress,showkeys,showpacs, 12pt]{revtex4-1}
\usepackage{amsmath,amssymb,amsthm,latexsym,bbm,calc}
\usepackage{graphicx,color}
\usepackage[bookmarks=false]{hyperref}
\newcommand{\be}{\begin{equation}}
\newcommand{\ee}{\end{equation}}
\newcommand{\beq}{\begin{equation}}
\newcommand{\eeq}{\end{equation}}
\newcommand{\bes}{\begin{eqnarray}}
\newcommand{\ees}{\end{eqnarray}}
\newcommand{\bqa}{\begin{eqnarray}}
\newcommand{\eqa}{\end{eqnarray}}
\newcommand{\bea}{\begin{eqnarray}}
\newcommand{\eea}{\end{eqnarray}}

\newcommand{\R}{\mathds{R}}
\newcommand{\C}{\mathds{C}}

\newcommand{\nn}{\nonumber}

\renewcommand{\R}{\mathbbm{R}}
\renewcommand{\C}{\mathbbm{C}}

\newcommand{\su}{\mathfrak{su}}

\newcommand{\iso}{\mathfrak{iso}}

\newtheorem{definition}{Definition}

\newcommand{\cA}{{\cal A}}
\newcommand{\tcA}{\tilde {\cal A}}

\newcommand{\cF}{{\cal F}}
\newcommand{\cC}{{\cal C}}
\newcommand{\cT}{{\cal T}}
\newcommand{\cG}{{\cal G}}

\newcommand{\cJ}{{\cal J}}

\newcommand{\cH}{{\cal H}}

\newcommand{\cS}{{\cal S}}
\newcommand{\cP}{{\cal P}}

\def\mg{{\mathfrak{g}}}

\def\mF{{\mathfrak{F}}}

\def\cs{{Lie \,\cG}}
\def\CS{{\mathcal{G}}}
\def\iso{{\mathfrak{iso}}}

\newcommand{\la}{\left\langle}
\newcommand{\ra}{\right\rangle}

\def\tell{{\tilde{\ell}}}
\def\tX{\tilde{X}}

\def\one{{\bf 1}}

\def\mhffp{{{h}_{cc'}}}

\def\mH{{{H}}}

\def\bfe{{\tilde {\bf e}}}
\def\bfx{{\bf y}}

\def\tH{{\tilde H}}

\def\ot{{\otimes}}
\def\dr{{\rightarrow}}
\def\mone{^{-1}}

\def\ov{\overline}

\def\demi{{\frac{1}{2}}}

\newcommand{\act}{{\triangleright}}
\newcommand{\wedcom }{\wedge}

\def\ovo{{\overline{\omega}}}
\def\ove{{\overline{\textbf{e}}}}

\DeclareMathOperator{\Tr}{Tr}

\DeclareMathOperator{\tr}{tr}

\DeclareMathOperator{\SU}{SU}
\DeclareMathOperator{\SL}{SL}

\def\poi#1{\{ #1 \}}

\newcommand{\rd}{\mathrm{d}}

\begin{document}

\title{\Large \bf Discretization of 3d gravity in different polarizations}

\author{{\bf Ma\"it\'e Dupuis}}\email{m2dupuis@uwaterloo.ca}
\affiliation{Department of Applied Mathematics, University of Waterloo, Waterloo, Ontario, Canada}
\affiliation{Perimeter Institute for Theoretical Physics, 31 Caroline St. N, ON N2L 2Y5, Waterloo,Canada}

\author{{\bf Laurent Freidel}}\email{lfreidel@perimeterinstitute.ca}
\affiliation{Perimeter Institute for Theoretical Physics, 31 Caroline St. N, ON N2L 2Y5, Waterloo,Canada}

\author{{\bf Florian Girelli}}\email{fgirelli@uwaterloo.ca}
\affiliation{Department of Applied Mathematics, University of Waterloo, Waterloo, Ontario, Canada}

\date{\small\today}

\begin{abstract}
\noindent We study the discretization of 3d gravity with $\Lambda=0$ following the loop quantum gravity framework. In the process, we realize that different choices of polarization are possible. This allows to introduce a new discretization based on the triad as opposed to the connection as in the standard loop quantum gravity framework. We also identify  the classical non-trivial symmetries of discrete gravity, namely  the Drinfeld double, given in terms of momentum maps. Another choice of polarization is given by the Chern-Simons formulation of gravity. Our framework also provides  a new discretization scheme of Chern-Simons, which keeps track of the link between the continuum variables and the discrete ones.
We show how the Poisson bracket we recover between the Chern-Simons holonomies allows to recover the Goldman bracket. There is also a transparent link between the discrete Chern-Simons formulation and the discretization of gravity based on the connection (loop gravity) or triad variables (dual loop gravity).  

\end{abstract}

\medskip

%\noindent  Pacs numbers: 02.10.Ox, 04.60.Gw, 05.40-a

\maketitle

\tableofcontents

%%%%%%%%%%%%%%%%%%%%%%%
\section*{Introduction}
%%%%%%%%%%%%%%%%%%%%%%%
Three-dimensional gravity is a topological theory that can be exactly quantized \cite{carlip}. Seen as a toy model for testing fundamental questions pertinent to four-dimensional quantum gravity, three-dimensional quantum gravity is also compelling for finding new approaches to the physically relevant four-dimensional theory. Three-dimensional gravity is as well appealing due to the several existing quantum models \cite{carlip}. For example, Loop Quantum Gravity and the combinatorial quantization formalism come from two different quantization and discretization processes.

Loop Quantum Gravity is a canonical quantization approach whose starting point is the first order gravity action, the so-called Palatini action \cite{rovelli}. In the quantization process, two different procedures are achieved: the discretization and the quantization itself. These two strokes are usually done in one, but let briefly describe them separately here. Discretization is done using a graph, embedded in the spatial surface, which in three dimensions can be chosen dual to a  triangulation. The continuous phase space variables, the connection and the triad, are respectively smeared along the edges of the graphs and along the corresponding dual edges. As a result, the discrete phase space is described in terms of the so-called holonomy-flux variables and their Poisson algebra is the the holonomy-flux algebra. The result of the second stroke is the definition of the quantum states of space, the so-called spin networks.  A key outcome of Loop Quantum Loop Quantum Gravity is the discreteness of  the geometric spatial operators. Loop Quantum Gravity is nevertheless not a discrete approximation of gravity. Indeed, one recovers the continuum picture by taking a projective limit on the Hilbert spaces \cite{thiemann}.  

The combinatorial quantization formalism is also a Hamiltonian quantization approach, which is based on the Chern-Simons action. The Chern-Simons action describes three-dimensional gravity but allows degenerate metrics as additional solutions. As previously mentioned, the discretization procedure is based on graphs embedded in the spatial surface. The  discretization of the classical Chern-Simons theory, or the Atiyah-Bott symplectic form \cite{Atiyah:1982fa}, has been first identified for (closed) holonomies, giving rise to the Goldman bracket \cite{goldman}. The Poisson algebra is not always well defined if one deals with open holonomies, but this problem was addressed by Fock and Rosly  \cite{Fock:1998nu}. They postulated  that some Poisson structures for the (open) holonomies, which allow to recover the moduli space with the right Poisson structure. They did not consider the link between the continuum variables Poisson bracket and their Poisson bracket between the holonomies. Later, Alekseev and Malkin proposed a change of coordinates in the phase space variables which made apparent nice structures such as the Drinfeld and Heinsenberg doubles \cite{Alekseev:1993rj}. Then, Alekseev, Grosse and Schomerus proposed  a direct quantization procedure of the Fock and Rosly phase space \cite{Alekseev:1994pa, Alekseev:1994au}.

 Although both continuum theories are describing gravity, the Loop gravity and Chern-Simons theories are difficult to compare (see nevertheless \cite{cat-kitaev}) at the discrete level and are written in terms of different mathematical structures at the quantum level.  \\
In this paper, we want to focus on the discretization step of both theories and clarify the link between the two approaches. In the Loop Gravity context, the discrete theory is interesting per se as it provides a way to truncate the theory into some approximation, getting in particular finite dimensional Hilbert spaces when the quantization is done. As in the quantum case, this truncation can be removed by considering the classical analogue of the projective limit. 

\medskip

Starting from the Palatini action for three dimensional gravity with a zero cosmological constant we reproduce the analysis of \cite{Freidel}, which was focusing on the four dimensional case. Identifying precisely the passage from continuum to discrete reveals that some \textit{choice of polarization} is actually made in the standard Loop Quantum Gravity approach. We discuss then the consequence of making \textit{other choices of polarization}.  We obtain three main results. \begin{itemize}
\item We show that a different choice of polarization allows us to define the corresponding phase space based on the metric (triad) picture. This new discrete framework where the vanishing of the curvature around each face of the triangulation is automatically implemented could be of interest to write a more geometrical formulation of gravity. 
 
\item Chern-Simons theory can be viewed as 3d gravity where  no specific choice of polarization is made.  We can also apply our "loopy" discretization scheme to Chern-Simons theory. The Poisson structure is well defined for any holonomy meaning that we have derived an alternative to  the usual Fock and Rosly regularization, while keeping a clear link  with the continuous Chern-Simons variables.

\item This metric kinematical phase space can be shown to be the dual of the usual Loop Quantum Gravity holonomy kinematical phase space through the notion of symplectic dual pairs. These two dual representations are unified within the Chern-Simons framework.  We illustrate in particular how different choices of polarization in the discretized Chern-Simons theory lead to the two dual discretized gravity pictures.
 
\end{itemize}

\medskip

The scheme of the paper goes as follows. In Section \ref{lqg and dual}, we describe the discretization procedure of the gravity phase space variables and the associated symplectic form following \cite{Freidel} for 3d gravity when $\Lambda=0$. We identify the dual representation of the Loop Quantum gravity phase space. In Section \ref{sec:CS}, we propose a new discretization of Chern-Simons theory following the Loop  Gravity discretization and we show that the Goldman brackets are recovered without introducing any ad-hoc regularization of the Poisson brackets. Finally in Section \ref{sec:CS-grav}, we show the link between the three different discrete pictures coming from either the Palatini action or the Chern-Simons action.

%
%%%%%%%%%%%%%%%%%%%%%%%
\section{3d LQG and its dual counterpart }\label{lqg and dual}
%%%%%%%%%%%%%%%%%%%%%%%

\subsection{Continuous phase space and constraints}

We consider  a principal G-bundle  over $M$, a 3d manifold (with no boundary).  We will consider in the following $\textrm{G}=\SU(2)$ or $ \SU(1,1)$. We note $\omega^A=\demi\epsilon^{ABC}\omega_{BC} $ its connection and $e_A$ the triad, which are both $\mg$ valued 1-form, with $\mg=\su(2)$ or $\su(1,1)$. 
The transformation properties are as follows
\beq\label{gauge transf}
\omega \dr \, \omega+  \rd \zeta+ [\omega\,,\, \zeta]  = \omega+\rd_\omega \zeta, \quad e\dr\, e+ [e,\zeta] \quad \textrm{with } \zeta \textrm{ a } \mg\textrm{-valued scalar}.
\eeq
The curvature of the connection is the $\mg$-valued 2-form $F=d\omega+ \omega\wedcom\omega$. 
Given the Lie algebra $\mg$, with generators $\sigma_A$, we write its   Killing form $\la,\ra$ as a normalised trace\footnote{In the case of $\su(2),$ or $\su(1,1)$, we can choose the generators to be antihermitian and to satisfy the algebra $[\sigma_A,\sigma_B]=\epsilon_{ABC} \sigma^C$  and the normalised trace to be 
related to the 2 dimensional trace by $\Tr(A) =-2 \tr(A)$. }
\be
\la \sigma_A, \sigma_B\ra= \Tr (\sigma_A\sigma_B) = \eta_{AB}.
\ee
The 3d gravity action with zero cosmological constant is given by  the \textit{BF} action \beq
\cS_{\rm grav}(e,\omega)=
- \int_M \la e \wedge F\ra= - \int_M  e^I \wedge F_I. 
\eeq
Capital indices are internal space indices, $I,J=1,2,3$. The choice of gauge group $\textrm{G}$ determines the signature of the spacetime under consideration. The equations of motion implement that  the connection should be torsionless and flat. 
\bes \label{EOM}
 \rd_\omega e= \rd e+ \omega \wedcom e=0, \quad \rd \omega + \omega\wedge \omega =0.
\ees
The \textit{BF} action is invariant under the gauge transformations \eqref{gauge transf}, but also the translation
\beq\label{translation infinit}
\omega\dr\, \omega, \quad e\dr\, e+\rd_\omega \phi, \quad \textrm{with } \phi \textrm{ a } \mg \textrm{ valued scalar},
\eeq
thanks to the Bianchi identity $d_\omega F=0$. 

\medskip

We assume that $M\sim \R\times \Sigma$ (with $\Sigma$ a smooth 2d manifold with no boundary) and use the coordinates $(t,x_1,x_2)$ for a point in $M$. We can then proceed to the Hamiltonian formulation, and identify the momentum variable which is of density weight 1. The natural choice is given by the  dyad of density weight 1, 
\be
\frac{\delta S_{grav}}{\delta \dot \omega_{aJ}}\equiv \bfe^a = \tilde \epsilon^{ab}e_b.
\ee
 Lower case indices are space indices, $a,b=1,2$ and $\tilde \epsilon^{ab}$ is the antisymmetric tensor of density weight 1 such that $\tilde \epsilon^{12}=1$. We  introduced $\delta $  a variational differential \cite{Donnelly:2016auv} acting on fields which squares to zero $\delta^2 =0$ and should not be confused with the space differential $\rd$.
As such the product $\delta A \,\delta B$ means the antisymmetric combination 
$\delta_{[1}A \,\delta_{2]} B$. We do not introduce a wedge notation for this skew symmetric product but we have to remember that the product of two variational forms is anti-commuting. 
We can then identify  the symplectic potential (Liouville form) $\Theta_{grav}^{\rm LQG}$. 
\be\label{liouville}
\Theta_{grav}^{\rm LQG}=\la  \bfe^a \, \delta \omega_a \ra=\la  \bfe\cdot \delta \omega \ra=\la  e\wedge \delta \omega \ra.
\ee
Dynamics is given in terms of a pair of constraints implementing that the spatial parts of the curvature or the torsion are zero.
\be
 F^I=\tilde\epsilon^{ab}F^I_{ab} =0, \quad T_J=(\partial_a \bfe_{J}^a+ {\epsilon_J}^{IK}\omega_{Ia}\bfe_{K}^a) = 0. 
\ee

Since a lot of attention will be given in the following to the symplectic potential, let us make some preliminary comments. The Liouville form we have obtained allows to identify the phase space variables and provides the symplectic form, which in turns provides the Poisson bracket. However there are in fact different possible choices of Liouville form, equivalent up to boundary terms. Let us recall the case of the particle. In this case, the standard Liouville term is $p\rd q$ with $p$ the momentum variable and $q$ the configuration variable. This Liouville term is equivalent up to boundary term to 
\be
- q\rd p, \quad\textrm{ or also } \demi(p\rd q - q\rd p).
\ee
Any of these Liouville forms lead to the same symplectic form (up to boundary terms). A similar ambiguity occurs in the gravity case. However since some of the phase space variables should be of density weight 1, we have different options. The LQG  case would consists in considering as the configuration variable the connection $\omega_a$, hence the Liouville term we have introduced earlier in \eqref{liouville}. 
The associated (non-zero) Poisson brackets would then read
\beq\label{symplectic 0}
\poi{\omega^I_a(x), \bfe^b_J(y)}= \delta^a_b\,\delta^I_J\,\delta^2(x-y) , \quad x,y\in \Sigma.
\eeq
Instead, we could consider as configuration  variable the dyad $e_a$, which would be more in the philosophy of the ADM formalism. This is in essence the dual picture to the LQG framework.  In this case the symplectic 2-form would then read\footnote{Recall that $\omega$ and $e$ are 1-forms and that we do an integration by parts.}  (up to boundary terms)
\be\label{adm liouville}
\Omega^{{\rm LQG}^*}_{grav}= \la \delta\tilde \omega^a \,\,\delta e_a \ra= \la \delta\tilde \omega \cdot \delta e \ra= \la \delta \omega \wedge \delta e \ra.
\ee
In this case, the momentum variables are given by $\tilde \omega^{Ia}(x) = \tilde \epsilon^{ab}\omega_b ^{I}(x)$. The Poisson bracket is then 
\be \label{dual bracket}
 \poi{e_{Jb}(x) , \tilde \omega^{Ia}(y) }= \delta^I_J\,\delta^a_b\,\delta^2(x-y).
\ee
\smallskip
Finally as we shall argue later, the symplectic 2-form 
\be\label{liouville grav-cs}
\Omega^{CS}_{grav}=\demi (\la  \delta\bfe \cdot \delta \omega \ra+ \la  \delta\tilde\omega \cdot \delta e \ra)=\demi (\la  \delta e \wedge \delta \omega \ra+ \la  \delta\omega \wedge \delta e \ra)
\ee will be related to the symplectic 2-form of Chern-Simons theory.
The next sections will consist in finding a consistent discretization  of these different symplectic 2-forms.

\medskip

The symmetries of the action given by the gauge transformations \eqref{gauge transf} or the translations \eqref{translation infinit} can also be realized in terms of the Poisson brackets. The momentum maps (that is the phase space functions that implement these symmetry transformations, see appendix \ref{symplectic}) are precisely these constraints. The curvature constraint $F$ implements the infinitesimal translation, whereas the  torsion constraint $T$ implements the infinitesimal gauge transformation. By considering the  smearing of the torsion and curvature $\cT=\int \zeta^K T_K$, $\cF=\int \phi_K F^K$ over the fields $N_K,\Lambda^K$, using the LQG phase space variables, we have that 
\bes
&&\poi{\bfe,\cT}= -[\bfe,\zeta], \quad \poi{\omega,\cT}=- \rd_\omega \zeta,\label{torsionb}\\
&& \poi{\bfe,\cF}= \, \widetilde{\rd_\omega   \phi} , \quad \poi{\omega,\cF}=0. \label{curvatureb}
\ees
The discretization scheme we will use should implement that the discretized constraints are the momentum maps implementing the discretized symmetries. 

\subsection{Towards the discretization of the gravity phase space }\label{lqg}
\begin{figure}[h]
\centering
\includegraphics[scale=.7]{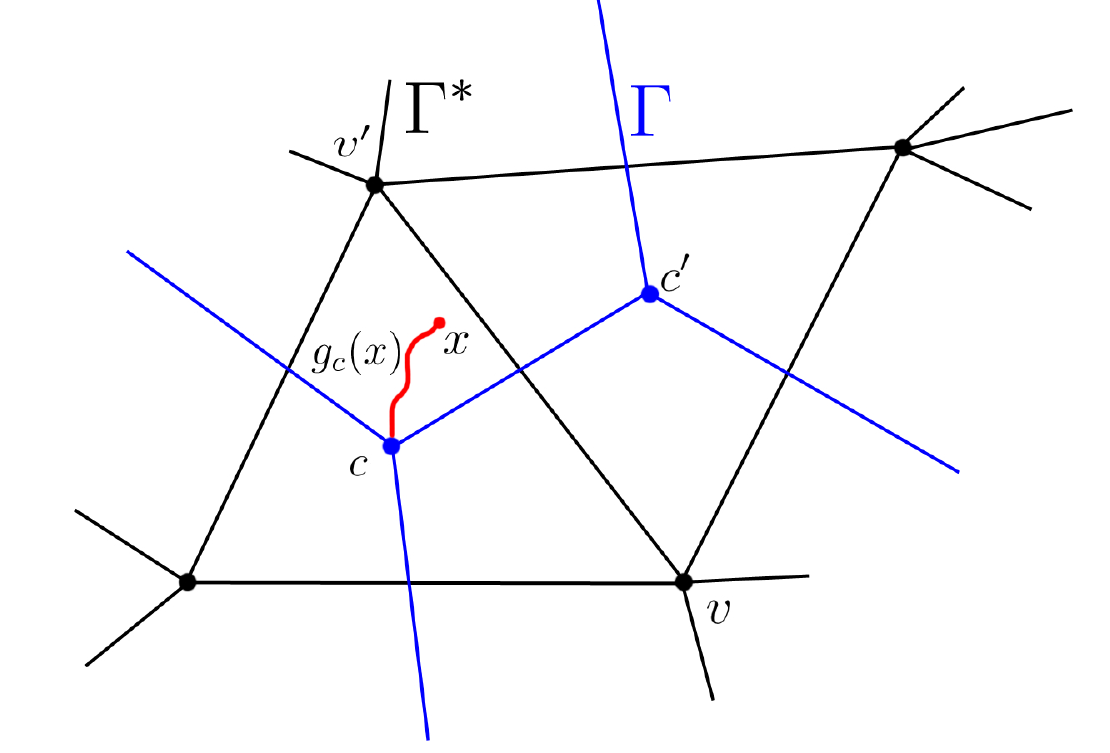}
\caption{Some components of the graphs $\Gamma$, $\Gamma^*$ and the holonomy $g_c(x)$. Curvature and torsion sit at the  vertices of $\Gamma^*$. The face $c^*$  and the face $c'^*$  share the edge $\ell^*=[vv']$, dual to the link $\ell=[cc']$.}\label{fig1}
\end{figure}

We intend to construct a discretization of the gravity symplectic form inspired by  previous works on discretizing gravity in order to get LQG \cite{Freidel}.
One first chooses a triangulation\footnote{We could more generally choose any cellular decomposition. We restrict to triangulations for the clarity of  exposition only.} $\Gamma^*$   of $\Sigma$ and we denote by $\Gamma$ the graph given by its one skeleton. In the following, we will denote the vertices of the triangulation $\Gamma^*$ by $v,v'$ and the oriented {\it edges} of $\Gamma^*$ by $\tilde{\ell}=[vv']$.
 We assume that in each  face $[v_1v_2v_3]$ of the triangulation a center point $c$ has been chosen, and we denote the duality between centers and triangles by $*$: $c^*= [v_1v_2v_3]$.
We connect the centers by {\it links} $\ell=[cc']$ and the graph made out of the centers and the links is denoted $\Gamma$ (see Fig. \ref{fig1}). This graph is dual to the triangulation graph $\Gamma^*$ and the duality between links and edges is written as
\be
\ell^*=[cc']^* =[vv']=\tilde \ell \quad  {\rm{if}} \quad [vv'] = c^* \cap c'^*.
\ee 
The duality is between oriented links and oriented edges. The orientation of the edge is chosen to be obtained from the orientation of the link by a counterclockwise rotation (see Fig. \ref{fig1}).

\medskip

We intend to discretize the phase space variables $(\omega,\bfe)$ in which $\omega$ is the configuration variable or $(\tilde \omega,e)$ in which $e$ is the configuration variable. There is a priori no recipe on  how the density weight  1  variables should be discretized. So instead we are going to discretize the 1-forms $(\omega,e)$ and see at the end that our discretization naturally provides a discretization of these   density weight  1 variables.

\medskip

 We  discretize the 1-form variables $(\omega,e)$ by assuming that the curvature $F(\omega) $ and torsion $T$, if any\footnote{We leave open the possibility that there is some by considering the dynamics later  or if we are introducing  particles.},  are concentrated 
 on the  vertices of $\Gamma^*$.
 This means that in the interior of each triangle $c^*$ we have $d_\omega e= 0 = d_\omega \omega$, that is the connection $\omega$ is flat and torsionless
 inside the cell $c^*$. These equations can be solved easily inside each triangle $c^*$ in terms of a group element $g_{c }(x)$, normalised to $g_c(c)=1$.
 This group element represents the {\it holonomy} of the flat connection from the center $c$ to the point $x\in c^*$ as illustrated in Fig. \ref{fig1}.
 The solution for the connection simply reads for $x\in c^*$
  \beq\label{def A e}
\omega (x)\equiv  (g_{c}^{-1} \rd g_{c})(x). 
\eeq
Given this parametrisation of $\omega$, the zero torsion condition implies 
that the combination $(g_c  e  g_c^{-1})(x) $  is closed hence exact on $c^*$.
Therefore we introduce the Lie algebra valued function $\bfx_c(x)$ on $c^*$ which solves the torsion condition as 
\be\label{def e}
e (x) \equiv \left(g_{c} \mone \rd\bfx_c  g_c\right)(x), \qquad x\in c^*.
\ee
The next step is to determine the discretization of the symplectic form behind the Poisson bracket \eqref{symplectic 0}.  Let us consider the variations of $\omega$ and $e$, in each triangle $c^*$, from their definitions in \eqref{def A e}.  A key identity that we will repeatedly use\footnote{ An explicit derivation reads 
\bes
\delta (h\mone \,  \rd h )=   h\mone \, \rd \delta h + \delta h\mone \, \rd h  
= h\mone\, \left(   (d \delta h) h\mone  -(\delta h   h\mone)\,(\rd h h\mone)\right) h= 
h\mone\, \rd \left(\delta h   h\mone\right) h.
\ees} is that 
\be {
\delta (g\mone \,  \rd g ) = g\mone\, \rd \left(\delta g   g\mone\right) g.}\label{gvar}
\ee
We also need to extract the variation of the electric field 
$e =   (g\mone \, \rd\bfx\,  g)$. In order to do so we similarly establish that
\be {
 \delta  (g\mone \, \rd\bfx\,  g) = 
 g \mone \, \left( \rd \delta \bfx \, +[ \rd\bfx , \delta g  g \mone]\right) g.}\label{yvar}
\ee 
Now that we have a simple expression of the fields inside each triangle $c^*$,
we can decompose the full discretized symplectic structure as a sum
$\Omega =\sum_c \Omega_c$, and 
  $\Omega_c$  is defined by  
\be 
\Omega_c=\int_{c^*}\Omega_{\rm grav}= \int_{c^*}  \la    \delta \omega {\wedge}  \delta e  \ra .
\ee
 
 From the value of $\delta \omega$ and $\delta e $ in $c^*$, we have 
\bes
\Omega_c
&=&  \int_{c^*} \la   \rd \left( \delta g_c g_c\mone\right){\wedge}   \left(\delta \rd \bfx_c +    [\rd\bfx_c ,  \delta g_c \, g_c\mone]\right)   \ra
= \int_{c^*}  \delta \la \rd\left( \delta g_c g_c\mone\right){\wedge}\,   \rd \bfx_c   \ra . \label{bou2}\label{bou0} \label{bou1}
\ees
The main point is that the integrand is an exact two form, it can therefore be entirely evaluated in term of its boundary contribution.
Also we remark that since $\bfx$ and $g$ enter asymmetrically there are two different ways to integrate this form. In other words  we have 
\bes 
 \Omega_c &=&   \int_{\partial c^*}    \delta\la \left( \delta g_c g_c\mone\right) \, \rd  \bfx_c  \ra \label{decomp} \\ 
 &=&  -  \int_{\partial c^*}\delta\la \rd \left( \delta g_c g_c\mone\right) \,     \bfx_c \ra \label{decomp 2}.
\ees
These two choices can be seen as the two natural choices of phase space variables $(\omega, \bfe)$ or $(\tilde \omega, e)$ respectively . Indeed  the tilde variable in which the  $\epsilon$ of weight density 1 sits, indicates what variable will be discretized on the dual of $\Gamma$. It can be the flux $\bfe$ , as in the Loop polarisation, or the holonomy in what we will call the dual Loop polarisation.

\subsection{Loop gravity phase space} \label{LQG phase space}
We now work with the loop polarisation $(\omega,\bfe)$ represented by the choice  (\ref{decomp}).
Since we have localised the symplectic structure on the boundary of the triangles we can equivalently write the total symplectic structure as a sum of contribution associated with each link $[cc']$ of $\Gamma$  as 
\be
\Omega^{\mathrm{LQG}} = \sum_{[cc'] \in \Gamma}   \Omega_{cc'},
\ee 
where the contribution from each link $\ell=[cc']$ is given by  contributions from $c^*$ and  from\footnote{Note that the face $c'^*$ contributes with a minus sign due to the opposite orientation.} $c'^*$. The edge shared by the faces $c^*$ and $c'^*$ is  $\tell=[vv']$, see Fig \ref{fig1}.  
\be\label{Symp1}
 \Omega_{cc'} = \delta  \int_{\tell }\left( \la  \left( \delta g_c g_c\mone\right)  \rd \bfx_c  \ra   - \la\left( \delta g_{c'} g_{c'}\mone\right) \rd \bfx_{c'}   \ra\right).
\ee
We  now look at the matching condition across the edge $[vv']=\tell$. Demanding the continuity of the connection across the edge implies that 
\be
 g_{c}^{-1} \rd g_{c}(x) = \omega(x)= g_{c'}^{-1} \rd g_{c'}(x),\qquad x\in [vv'].
\ee
This condition means that there is a {\it constant } group element $h_{cc'}$ that relates both frames 
\be\label{gframe}
g_{c'}(x) = h_{c'c} g_{c}(x),\qquad x\in [vv']. 
\ee
$h_{cc'}$ represents the {\it holonomy} of the flat connection along the edge $[cc']$ of $\Gamma$. In the following we use that $ h_{cc'}^{-1}= h_{c'c}$. $h_{cc'}$ represents the usual group variables of loop gravity, in which $\Gamma$ is the support of  the spin network.
\begin{figure}[h]
\centering
\includegraphics[scale=.7]{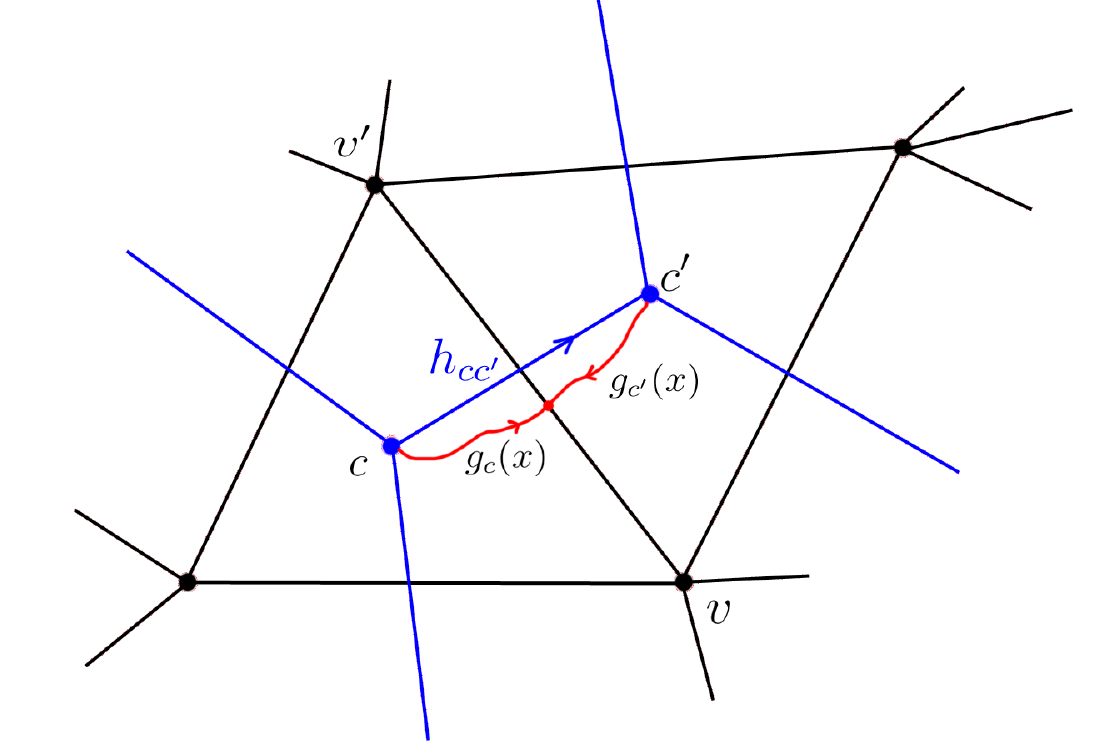}
\caption{The constant holonomy $h_{cc'}$ connects the frames of the two different faces. It is the standard LQG holonomy decorating the spin network $\Gamma$. }
\label{fig23}
\end{figure}

The frame field $e(x)$ is also continuous, hence we can relate the  frames from different faces.  Consider $x\in \tell$ then we have
\be
 g_c\mone \rd\bfx_c g_c = e =  g_{c'}\mone \rd\bfx_{c'} g_{c'}. 
 \ee
Hence we deduce that for $x\in \tell$,  we have 
$
\rd\bfx_{c'} = \mhffp \mone \, \rd\bfx_c\,  \mhffp .
$
This can be integrated out and we conclude that there are elements $x_{cc'}\in \mg$ such that 
\be \label{xframe}
\bfx_{c'}= \mhffp \mone (\bfx_c + x_{cc'}) \mhffp .
\ee
$x_{cc'}$ represents the translational holonomy.
In other words when going from $c$ to $c'$ the frame is rotated by $h_{cc'}$ but also translated by $x_{cc'}$. The combination 
$(h_{cc'}, x_{cc'})$ represents a Poincar\'e transformation that maps the flat chart around $c$ to the flat chart around $c'$.

\medskip
In order to evaluate the symplectic structure $\Omega_{cc'}$ we consider the
variation of the frame relation (\ref{gframe}) and get that 
\be\label{deltagc}
\delta g_{c'} g_{c'}\mone = h_{cc'}^{-1} \left(\delta g_c g_c^{-1} - \delta h_{cc'}h_{cc'}^{-1}\right)h_{cc'}.
\ee 
This together with (\ref{xframe}) leads us to the simple expression 
\be\label{Om1}
\Omega_{cc'}=   \delta
\la\left( \delta h_{cc'}h_{cc'}^{-1}\right)|  \tilde{X}^c_{\tell}   \ra,
\ee
where we have introduce the {\it flux}  vector $\tilde{X}_{\tell} \in \mg $ based at  $c$, and defined as 
\be\label{flux1}
\tilde{X}_{\tell}^c = \int_{\tell} \rd\bfx_c = \int_{\tell} \left(g_c \,e \, g_c\mone\right)\equiv \int_{\ell}\left(g_c \,\bfe \, g_c\mone\right).
\ee
Such flux is the natural candidate to encode the discretization of the density weight 1 vector $\bfe^a$. The presence of the 2d Levi-Civita tensor can be traced back to the fact that $\bfe$ is discretized on the dual of $\ell$, on which the connection is discretized through the holonomy $h_\ell$.  

\medskip

The discretized variables are therefore 
\be\label{LQG variables}
(\tilde X_{\tilde \ell}^c,  h^v_{\ell}= g_{cv}g_{c'v}\mone) .
\ee 
We picked the point $v$ in $\tell$ to define $h_{\ell}$ to emphasize the symmetry between these variables and the dual LQG variables \eqref{dual LQG variables}.
In the following, when dealing with the usual LQG variables, we will drop the upper indices $c$ and $v$ to avoid cluttered notations.

\smallskip

The symplectic form  associated to the link $\ell= [cc']$  takes then the following shape 
\bes
&&  \Omega^{LQG}_{cc'}= \Tr \left(  (\delta h_\ell h_\ell\mone)\delta \tilde{X}_{{\tell}}  +   (\delta h_\ell h_\ell^{-1}) ( \delta h_\ell h_\ell^{-1}) \tilde{X}_{\tell} \right).
\label{lqg 2-form}
\ees 
We have recovered the standard symplectic form associated to $T^*\textrm{G}$. Extending the construction to all of the edges of  $\Gamma$, we can decorate these edges with the phase space $T^*\textrm{G}$. The fluxes $\tilde X_{\tilde \ell}$ sit at the nodes of $\Gamma$ but depend on the edges $\tell$ of the graph $\Gamma^*$, whereas the holonomies $h_\ell$ decorate the links $\ell$ of $\Gamma$. This is the definition of the building blocks of the LQG phase space.

Note that we can invert the symplectic form to recover  the standard Poisson bracket on $T^*\textrm{G}$ (see for example \cite{simone} for the details of the calculations).
\be\label{poisson lqg}
\poi{\tX^A_{\tell},\tX^B_{\tell}} = \epsilon^{AB}_C \tX^C_{\tell}, \quad \poi{\tX^A_{\tell},h_{\ell}}= \sigma^A\, h_\ell, \quad \poi{h_\ell,h_{\ell}}=0.
\ee

It is important to recognize that the data $(\tilde{X}_{\tell}=\tilde X_{[vv']} ,h_{\ell})$ is not free.
It does satisfy a discrete version of the Gauss law. This constraint follows from the continuous Gauss identity
 \be\label{Gauss}
 \cJ_c\equiv \tilde{X}_{[v_1v_2]}+ \tilde{X}_{[v_2v_3]} + \tilde{X}_{[v_3v_1]}= \int_{\partial c^*}   \rd\bfx_c  
 =\int_{c^*}  \rd \left(\rd \bfx_c  \right) =  \int_{c^*} g\mone_c (\rd_\omega e) g_c
 =0. 
\ee
which is essentially the sum over the 3 boundary edges of the triangle $c^*$ of $\Gamma^*$.

We see from this formula that $ \cJ_c$ computes the 
violation of the torsion condition in the triangle $c^*$. Since we have assumed that it vanishes inside each triangle, the sum vanishes and we recovered the standard discretized Gauss law. 

\medskip 
Since torsion is the momentum map at the infinitesimal level implementing the infinitesimal gauge transformation \eqref{torsionb}, and that  $\cJ_c$ can be seen as the discretization of torsion on $c^*$, it is natural to expect that $\cJ_c$  encodes a local  ${\rm{G}}$ transformation at the trivalent vertex $c$ of $\Gamma$. As discussed in  Appendix \ref{symplectic}, $\cJ_c$ is the momentum map for the three copies of $T^*\textrm{G}$. Denoting $\cJ_c(\alpha) = \alpha^c_A\cJ_c^A$  we have 
\bes
\delta^c_\alpha \tX_{\tell}=\poi{\tX_{\tell}, \cJ_c(\alpha)  }, \quad \delta^c_\alpha h_{\ell} =\poi{h_{\ell}, \cJ_c(\alpha)  }.
\ees
As such, using \eqref{poisson lqg}, it implements the  transformation
 $\delta^c_\alpha$ with  $\alpha^{c}\in \mg$  at the nodes $c$ of $\Gamma$. 
\be \label{center transf}
\delta^c_\alpha \tX_{\tell}=[\alpha_c, \tX_{\tell}], \quad \delta^c_\alpha h_{\ell} = -\alpha_c h_{\ell}. 
\ee 

\medskip

Following the Marsden-Weinstein theorem \cite{Alekseev:yg, lu},  we recover the usual kinematical LQG phase space as the double quotient:
\be\label{P1}
\cP_{\rm LQG}^{kin} =(\times_{\ell\in \Gamma} T_\ell^*{\rm{G}})/\!/(\times_{c\in \Gamma} {\cJ}_c ),
\ee
where the double quotient denotes the symplectic reduction by the constraint (\ref{Gauss}).
The decorated graph $\Gamma$ is the classical analogue of the spin network. 

\begin{figure}[h]
\centering
\includegraphics[scale=.6]{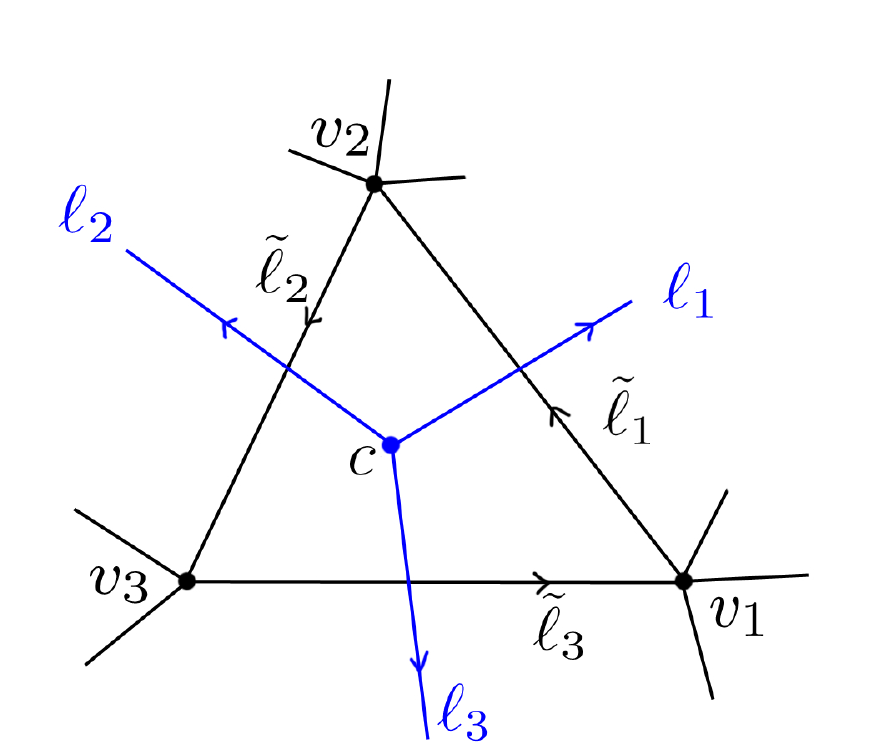}
\caption{The geometry of the triangle $c^*$ can be recovered from the vectors $\tX^c_\tell$. }
\label{triangle1}
\end{figure}

\medskip
The kinematical observables are functions built from the fluxes $\tilde X_{\tilde \ell}$ and holonomies $h_\ell$, such that they are invariant under the  transformations \eqref{center transf}. Typically such observables are then the Wilson loops $\Tr \prod_{\ell\in \lambda} h_\ell$, where $\lambda$ is a loop in $\Gamma$  or the scalar quantities built out from the fluxes. For example, the (kinematical) observables associated to the  (triangle edge) length, (triangle) angle,  (triangle) area, respectively $L_{v_1v_2}, \theta_{v_1}, A_{v_1v_2v_3}$ are given by 
\be
L^2_{v_1v_2}= |\tX_{[v_1v_2]}|, \quad \cos \theta_{v_1} = \frac{\tX_{[v_1v_2]}\cdot \tX_{[v_1v_3]}}{L_{v_1v_2}L_{v_1v_3}}, \quad A_{v_1v_2v_3}= \demi  |\tX_{[v_1v_2]}\wedge \tX_{[v_1v_3]}|.
\ee
These observables allow to reconstruct the triangle geometry dual to the vertex $c$, see Fig. \ref{triangle1}. 

\medskip

To have the full description of 3d gravity at the discrete level, we just need to introduce the discretized version of the flatness constraint (considering  the pure gravity case with no particles). This constraint is discretized by requiring that all the holonomies along the loops $\lambda$ in $\Gamma$ are flat.
\be
\cG_{\lambda}=\prod_{\ell\in\lambda} h_{\ell} =1.
\ee
This set of constraints together with the set of constraints generated by $\cJ_c$ form a first class system of constraints, just like in the continuum case.  

\medskip

The set of constraints $\{\cG_\lambda\}_{\lambda \textrm{ loops in } \Gamma}$  also implements some symmetry transformations. Note however that since we are dealing with elements in a non-abelian group, this constraint can be seen as a \textit{non-abelian} momentum map. 
We consider therefore the  symmetry action, labelled by the vertex $v$ and $\beta_v= \beta_v^A \sigma_A\in \su(2)$, given by
\bes 
&&\delta^v_\beta \tilde X_{\tell} ^B
\equiv \la \cG_{\lambda}\mone \poi{\tilde X^B_{\tell},\cG_{\lambda}},\beta_v^A \sigma_A\ra
, \quad \delta^v_\beta h_{\ell}
\equiv \la \cG_{\lambda}\mone \poi{h_\ell, \cG_{\lambda}},\beta_v^A \sigma_A\ra, \label{translation discrete}
\ees
Let us consider the loop $\lambda=(c_1c_2..c_i..c_nc_1)$ surrounding a vertex $v\in \Gamma^*$, with initial point $c_1$ and for $i=1,..n$, we define $\ell_i=[c_ic_{i+1}]$ with $[c_nc_{n+1}]\equiv[c_nc_{1}]$. The infinitesimal action $\delta^v_\beta$  is then explicitly choosing the ordering $\cG_{\lambda}=h_{\ell_{n}}\cdots h_{\ell_1}$, we have 
\bes
&&\delta^v_\beta \tilde X_{\tell_1}= h_{\ell_1}\mone \,\beta_v\,h_{\ell_1}, \quad \delta^v_\beta \tilde X_{\tell_2}= h_{\ell_2}\mone h_{\ell_1}\mone \,\beta_v\,h_{\ell_1} h_{l_2},\quad\cdots ,\quad \delta^v_\beta \tilde X_{\tell_i}= H_i\mone\,\beta_v \,H_i, \label{vertex transf}
\ees 
where we have defined the partial holonomies 
$H_i\equiv h_{\ell_1}h_{\ell_2}..h_{\ell_{i}}$
while it leaves the holonomies invariant
$  \delta^v_\beta h_{\ell_i}=0$.
We  recognize  the discretized action of the translations \eqref{translation infinit}. This transformation seems to depend on the choice of initial point one starts with. However the condition $ \prod_{\ell\in\lambda} h_{\ell} =1$ means that the different transformations generated by the different choices of initial point are all equivalent and related by a redefinition of the gauge parameter.

\smallskip

A momentum map with value in a non-abelian group indicates the presence of a  symmetry group (here the translations $\R^3$) equipped with a non-trivial Poisson structure, hence a non-trivial Poisson-Lie group. This is reviewed in the appendix \ref{symplectic}. At the quantum level, this leads to the notion of quantum group (here the Drinfeld double) as symmetry group.

\medskip 

The Marsden-Weinstein theorem has been generalized to the case where the symmetries are Poisson-Lie group symmetries \cite{lu}. Hence we can consider the symplectic reduction of the LQG kinematical phase with the symmetry action, acting at the vertices of the graph $\Gamma^*$ (vertices dual to the loops $\lambda\in\Gamma$). 
\be\label{P2}
 \cP^{phys}_{\rm LQG} = \cP_{\rm LQG} /\!/(\times_{{\lambda}\in \Gamma} {\cG}_{\lambda} ).
\ee

\medskip

The physical observables, that is the functions over copies of $T^*\SU(2)$ invariant under the symmetries spanned by $\cJ_c$ and $\cG_\ell$, have been discussed in \cite{ale-karim}.

\subsection{A dual representation of the loop gravity phase space}\label{LQG* phase space}
 In the previous section we have shown how the continuum symplectic  structure reduces to the discrete loop gravity one. In order to do so we have chosen the first decomposition in (\ref{decomp}). This is the polarisation were wave functions are functional of the connection. We now investigate what happens if we choose the second one, that is the geometrical polarisation where wave functions are functions of the frame $e$. In this case we can write the edge symplectic structure as\footnote{This evaluation can be compared to \eqref{Symp1}. Once again the face $c'^*$ has the opposite orientation and we have implemented the minus sign coming from the integration by part in \eqref{decomp}. } 
\be\label{Symp2}
 \tilde{\Omega}_{cc'} = \delta  \int_{\tell }\left(  \la \rd \left( \delta g_{c'} g_{c'}\mone\right)  \bfx_{c'}  \ra -\la \rd \left( \delta g_c g_c\mone\right)   \bfx_c  \ra   \right).
\ee
This term  differs from (\ref{Symp1}) by a boundary term, which disappears when we sum over all links. Indeed we have that 
\be
 \tilde{\Omega}_{cc'}  ={\Omega}_{cc'}  
 + \delta  \int_{\tell }\rd \left(  \la  \left( \delta g_{c'} g_{c'}\mone\right)  \bfx_{c'}  \ra -\la  \left( \delta g_c g_c\mone\right)   \bfx_c  \ra   \right).
\ee
Using (\ref{xframe}) and (\ref{deltagc}) we can evaluate it as 
\be 
\tilde{\Omega}_{cc'} = \delta \int_{\tell } \la \rd \left( \delta g_{c} g_{c}\mone\right)  x_{cc'}  \ra.
\ee
Using the fact that the edge dual to the link $\ell$ is given by  $\tell=[vv']$, 
and defining $g_{cv}:=g_c(v)$ we get 
\be 
\tilde{\Omega}_{cc'} = \delta  \la \left[\left( \delta g_{cv'} g_{cv'}\mone\right) - \left( \delta g_{cv} g_{cv}\mone\right)  \right] x_{cc'}  \ra.
\ee
If one introduces the holonomy from $v$ to $v'$  inside $c$:
\be\label{hcvv}
\tilde{h}_{\tell}^c=\tilde{h}^c_{vv'} \equiv g_{cv}^{-1}g_{cv'},  
\ee
we  get an equivalent but  more familiar  expression dual to (\ref{Om1})
\be \label{again}
 \tilde{\Omega}_{cc'} =  \delta \la  \delta \tilde{h}_{\tell}^c (\tilde{h}_{\tell}^c)^{-1}|  X_{\ell}^v  \ra,
\ee
where we have defined 
\be\label{flux2}
X_{\ell}^v\equiv  (g_{cv}^{-1} x_{cc'}g_{cv}).
\ee
$x_{cc'}$ is a translational monodromy based  at $c$, the connectors $g_{cv}$ map it onto a field based at $v$. 
It is important to note that from the definition we have the relation
\be\label{inverse flux dual}
X_{-\ell}^v= X_{[c'c]}^v= - (\tilde{h}^c_{vv'})^{-1} X_{[cc']}^v \tilde{h}^c_{vv'}= - (\tilde{h}^c_{\tell}) \mone X_{\ell}^v (\tilde{h}^c_{\tell})\mone.
\ee
To summarize, the variables for the dual LQG formulation are given by 
\be\label{dual LQG variables}
(X_{\ell}^v\equiv  (g_{cv}^{-1} x_{cc'}g_{cv}), \,\,\tilde{h}_{\tell}^c= g_{cv}^{-1}g_{cv'})
\ee
These variables are based at $v$ and can be seen as the dual picture to the standard LQG variables given in \eqref{LQG variables}. The dual fluxes are depending on $\Gamma$ whereas the holonomies depend on the triangulation $\Gamma^*$.

\medskip

Our construction provides a candidate for the discretization of the density weight 1 variable $\tilde \omega$.
\be\label{disc tilde omega}
\tilde h^c_\tell = P\exp\left(\int_\tell  \omega \right) \equiv P\exp\left(\int_\ell \tilde \omega \right).
\ee
Once again the presence of the Levi-Civita tensor implied that the vector $\tilde \omega^a$ was discretized over the dual space of where the 1-form $e$ was discretized (into $x_{\ell}$).

\medskip

\begin{figure}[h]
\centering
\includegraphics[scale=.7]{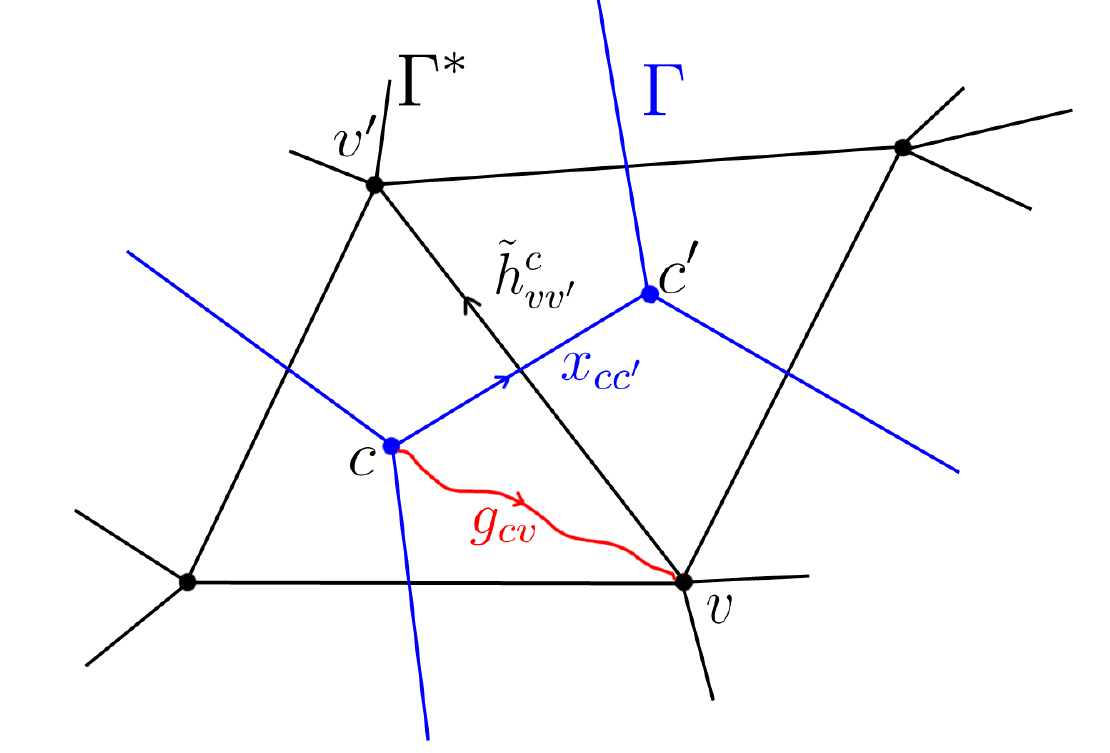}
\caption{The constant (abelian) holonomy $x_{cc'}$ connects the frames of the two different faces as a translation. The holonomy $\tilde{h}^c_{vv'}$ lives on the face $c$ and connects the vertices of $\Gamma^*$. We have now a dual picture of the standard LQG picture, based on $\Gamma^*$.}
\label{hcvvfig}
\end{figure}

The symplectic structure \eqref{again} is again the symplectic structure of $T^*\rm{G}$ 
for the pair $(X_{\ell}, \tilde{h}_{\tell}^c)$ but based on the dual graph  
$\Gamma^*$ instead of $\Gamma$. The associated Poisson bracket is once again 
\be\label{poisson lqg dual}
\poi{X^A_{\ell},X^B_{\ell}} = \epsilon^{AB}_C X^C_{\ell}, \quad \poi{X^A_{\ell},\tilde h^c_{\tell}}= \sigma^A\,\tilde h_{\tell}, \quad \poi{\tilde h^c_{\tell},\tilde h^c_{\tell}}=0.
\ee
We emphasize again that the holonomy is depending on the face $c$ from its definition \eqref{hcvv}, see Fig \ref{hcvvfig}.

\medskip

Just like in the standard LQG case, the variables we have integrated, namely here the holonomies, are not all independent. There is still a constraint naturally present due to the integration around $c^*$.
\be\label{Gaussg}
\cG_c\equiv \tilde{h}^c_{v_1v_2}\tilde{h}^c_{v_2v_3}\tilde{h}^c_{v_3v_1}=1.
\ee
where $(v_1,v_2,v_3)$ are the three vertices of the triangle $c^*$.
These constraints are the group analog  of the Gauss constraints (\ref{Gauss}) but now they express the vanishing of the curvature inside $c^*$.

\medskip

Once again, we can expect that the discretization of the curvature  generates a momentum map $\cG_c$, with value in a non-abelian group, spanning the translations. We consider therefore the infinitesimal symmetry action, labelled by the node $c$ dual to the triangle $c^*=[v_1v_2v_3]$ and $\beta_c\in\R^3\sim \su(2)$, given by
\bes 
&&\tilde\delta^c_\beta  X^B_{\ell} 
\equiv \la \cG_c \mone \poi{ X^B_{\ell},\cG_c},\beta_c^A \sigma_A\ra
, \quad \tilde\delta^c_\beta \tilde{h}^c_{\tell}
\equiv \la \cG_c\mone \poi{\tilde{h}^c_{\tell}, \cG_c},\beta_c^A \sigma_A\ra. \label{translation discrete dual}
\ees
This is to compare with the transformations generated by the LQG discrete flatness constraint \eqref{translation discrete}. Explicitly, these transformations gives, choosing the ordering $\cG_c=\tilde h^c_{\tell_{1}}..\tilde h^c_{\tell_3}$ and the orientation as in Fig. \ref{triangle1},
\bes\label{trans dual}
&&\tilde\delta^c_{\beta} X_{\ell_3}^{v_3} = (\tilde h^c_{\tell_3})\mone \beta_c \tilde h^c_{\tell_3}, \quad
\tilde\delta^c_{\beta} X^{v_2}_{\ell_2} = (\tilde h^c_{\tell_2}\tilde h^c_{\tell_3})^{-1} \beta_c\tilde h^c_{\tell_2}\tilde h^c_{\tell_3},
\quad 
\tilde\delta^c_{\beta} X^{v_1}_{\ell_1} = \cG_c \mone \beta_c \cG_c=\beta  \\
&&\tilde\delta^c_{\beta} \tilde{h}^c_{\tell_i} =0. \nn
\ees
This transformation seems to depend on the choice of vertex one starts with, and accordingly it looks like one should have three transformations per center. However the condition $\cG_c=\tilde h^c_{\tell_{1}}\tilde h^c_{\tell_2} \tilde h^c_{\tell_3}=1 $  means that these three transformations are all equivalent and related by a 
redefinition of the gauge parameter $\beta_c \to \tilde{h}^c_{\tell_1} \beta_c (\tilde{h}^c_{\tell_1})^{-1} \to
(\tilde h^c_{\tell_1}\tilde h^c_{\tell_2})^{-1} \beta_c\tilde h^c_{\tell_1}\tilde h^c_{\tell_2} $.

\medskip 

We define now the phase space of the dual LQG picture by considering the double quotient, 
\be\label{P3}
\cP^{kin}_{\rm LQG^*} =(\times_{\tell\in \Gamma^*} T_\tell^*{\rm{G}})/\!/(\times_{c\in \Gamma} {\cG}_c ),
\ee
where the double quotient denotes the symplectic reduction by the constraint (\ref{Gaussg}), thanks to the Marsden-Weinstein theorem generalized to the  non-abelian momentum map case \cite{Alekseev:yg}. Upon quantization, we expect to choose to construct our kinematical Hilbert space on the functions of $X_\ell$ together with the flatness constraint which implement a translation invariance at the centers of $\Gamma$. Hence we expect to recover some spin networks based on the translational group. Our phase space is the classical analogue of such spin (or "momentum") network. 
We note that in the dual picture we implement at the kinematical level the non-trivial Poisson Lie symmetry. Hence at the quantum level, we shall expect to deal with representations of a quantum group (the Drinfeld double).

\medskip

The kinematical observables invariant under the gauge transformations \eqref{trans dual} are the group elements themselves, but also  some new observables located at vertices $v$.

 \begin{figure}[h]
\centering
\includegraphics[scale=.5]{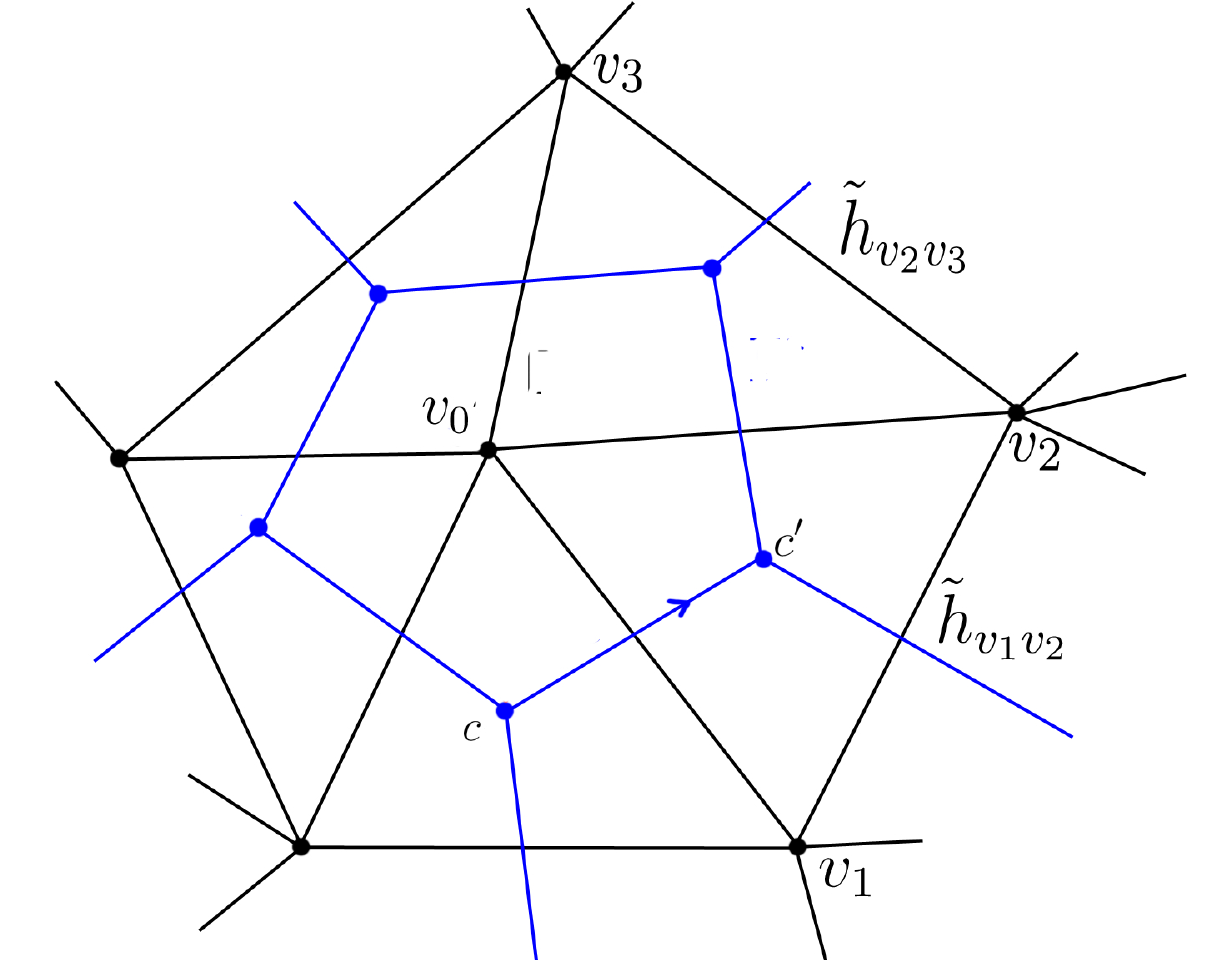}
\caption{We construct an observable associated to a vertex $v_0$.}
\label{new obs}
\end{figure}

Lets consider a vertex $v_0$ surrounded by $n $ vertices $(v_1,\cdots, v_n)$
ordered counterclockwise, 
and $n$ centers $(c_1, \cdots c_n)$, as in Fig \ref{new obs}. As earlier, we will note $\lambda$ the holonomies around the vertices $v$.
We chose the labels  such that $c_a^*=[v_0v_av_{a+1}]$, with $v_{n+1}=v_1$. 
For simplicity we denote $X_{[v_0v_{a}]^*}\equiv X_{a}$, and we notice that all the $X_a$ do sit at the vertex $v_0$.
Let us consider the triangles sharing the edges $[v_0v_1]$, see Fig \ref{new obs}. The flux $X_1$ gets  gauge transformed from the face $c^*_0$ and $c^*_1$. Note however that from the perspective of the face $c^*_1$, $X_1$ has an opposite orientation. Hence we should have in mind \eqref{inverse flux dual}.
\bes
\tilde\delta_{\beta_{c_0}}X_{1}^{v_0} = \beta_{c_0},\quad
\tilde\delta_{\beta_{c_1}} X_{1}^{v_0}= -\beta_{c_1}.
\ees
Repeating the process for each edge $[v_0v_a]$, we see that we can construct a kinematical observable (which commutes with $\cG_c$)
\be
\cJ_{v_0} = \sum_a X^{v_0}_a,
\ee
which computes at the discrete level the integral of the torsion  $\int_D \rd_\omega e$ around a  disk $D$ centred at $v_0$. This is the abelian holonomy around $v_0$. 

\medskip

Let us note $\alpha_{v} = \alpha_{vA} \sigma^A$.  $\cJ_v(\alpha_v) \equiv \alpha_{vA} \sum_{\ell\in\lambda} X_{\ell}^A $ is an abelian momentum map which  generates the following gauge transformations
\bes 
&&\tilde\delta^v_\alpha  X^B_{\ell} 
=   \poi{ X^B_{\ell},\cJ_v (\alpha_v)}= [\alpha_v, X^B_{\ell}]
, \quad \tilde\delta^v_\alpha \tilde{h}^c_{\tell}
=  \poi{\tilde{h}^c_{\tell},\cJ_v(\alpha_v)}= -\alpha_v \tilde{h}^c_{\tell}. \label{rotation discrete dual}
\ees

\medskip

The dynamics of the dual formulation of LQG will be given by the constraint $\cJ_v=0$. By construction, $\cG_c$ and $\cJ_v$ form a first class system (in fact they commute strongly, unlike the standard LQG case where they do commute weakly).

The physical phase space of the dual LQG formulation is obtained by   considering the symplectic reduction of the dual LQG kinematical phase by the symmetry action generated by $\cJ_v$, acting at the vertices of the graph $\Gamma^*$. 
\be\label{P4}
 \cP^{phys}_{\rm LQG^*} = \cP_{\rm LQG^*} /\!/(\times_{v\in \Gamma^*} {\cJ}_{v} ).
\ee

\section{A new discretization of Chern-Simons theory}\label{sec:CS}
To summarize the previous Section, we have seen that we have two  descriptions of the 
  gravity phase space, one based on the connection picture, the other based on a triad (metric) picture. We would like to see if we can strengthen the relations between the two pictures, by \textit{unifying} them. For this we can expect that the Chern-Simons approach might provide the key to this unification. Indeed, in the Chern-Simons approach to gravity,  the connection and the triad are put at the same level and unified through the Chern-Simons connection. 
  
We have seen that the dual loop gravity picture can be seen as  a different choice of polarization, having the triad as the configuration variable instead of the connection as in loop gravity. Another "choice" of polarization is to just remove the difference between configuration and momentum variables and work with these variables altogether. This is what the Chern-Simons connection does by unifying the triad and the connection. It provides a choice of coordinates on the gravity phase space agnostic in terms of what we call momentum or configuration.

We can therefore discretize the Chern-Simons theory along the same way we proceeded  for gravity. This will allow to simplify the relation between the (dual) loopy picture and Chern-Simons. As such we will provide a  discretization of Chern-Simons which will keep clear the link between the continuum picture and the discrete picture (which is not the case for the Fock-Rosly formalism \cite{Fock:1998nu}).  Our discretization scheme is also free of the issues that arise in a naive discretization (for example the Poisson bracket between holonomies sharing an initial point would diverge).  

To show that our discretization is consistent we will prove that we can recover the Goldman bracket \cite{goldman}. The discretization will be done for any group and in the next Section, we will focus on the Poincar\'e/Euclidian group cases to connect with gravity with zero cosmological constant.

\subsection{Chern-Simons theory}

We consider  a principal $\CS$-bundle  over $M$, a 3d manifold (without any boundary).   We note $ \cA$ its connection  which is a $\cs$-valued 1-form. We note $\la,\ra$ or $\eta^{\mu\nu}$  the invariant form. 
The gauge transformation properties and the curvature tensor are the standard ones 
\bes \label{action gauge}
&& \cA  \dr \cA+\rd_\cA \xi 
,  
\qquad
\mF=\rd\cA+ \cA{\wedge} \cA, \quad \mF\dr \mF+[\mF\,,\, \xi]\quad \textrm{with } \xi \textrm{ a } \cs \textrm{ valued scalar} .
\ees
The Chern-Simons action is given by 
\beq
\cS(\cA)= \int \demi\la \cA \wedcom \rd \cA\ra + \frac13 \la \cA\wedcom [\cA\wedcom\cA] \ra. 
\eeq
 The equations of motion implement that the connection $\cA$ is flat. 
\bes
 \mF =0.
  \ees

\medskip

Let us  assume now that $M\sim \R\times \Sigma$ (with $\Sigma$ a smooth 2d manifold with no boundary) and use the coordinates $(t,x_1,x_2)$ for a point in $M$. We can then proceed to the Hamiltonian formulation.
We identify the momentum variables to be the connection with density weight 1.
\be
\frac{\delta S_{\rm CS}}{\delta \dot \cA_{a\mu}}\equiv \tilde \cA^{a\mu} = \demi \tilde \epsilon^{ab}\cA_b^\mu.
\ee
Lower case indices of the beginning of the alphabet  are space indices, $a,b=1,2$, while greek indices are internal indices. The  symplectic 2-form is then 
\be
\Omega^{\rm CS}= \la \delta \tilde\cA\cdot\delta \cA\ra =\demi \la \delta \cA\wedge\delta \cA\ra ,
\ee
while the dynamics is given in terms of the constraint
\be
\tilde \epsilon^{ab}\mF^{\mu}_{ab}=0.
\eeq
The canonical Poisson bracket is obtained by inverting the symplectic form given by $\Omega^{\rm CS}=\delta \Theta^{\rm CS}$
\beq\label{symplectic CS}
\poi{\cA^{\mu a} (x),\tilde \cA_{\nu b } (y)}=  \,\delta ^a_b\delta^\mu_ \nu\delta^2(x-y), \quad x,y\in \Sigma.
\eeq
We note that  $\cA^{\mu a} (x)$ and $\tilde \cA_{\nu b} (y)$ are dual to each other. Hence upon discretization, we expect that the analogue of \eqref{symplectic CS} should be given in terms of holonomies living on dual edges.

Later on, we will also use  the following notation for the Poisson bracket\footnote{ We use the standard notation $\cA^{(1)}=\cA^\mu T_\mu\ot \one$ and  $\cA^{(2)}= \one \ot \cA^\mu T_\mu$ and similarly for $\tcA$. $T_\mu$ are the generators of $\cs$.}
\be\label{CS poisson}
\poi{\cA^{(1)} (x),\tilde \cA^{(2)} (y)}=  \delta^2(x-y) T^\mu\ot T_\mu , \quad x,y\in \Sigma.
\ee

\medskip 

A direct calculation shows that the smeared curvature  is the momentum map implementing the infinitesimal gauge transformations \eqref{action gauge}. For example
\bes
\poi{ \cA, \int \xi^\nu \mF_\nu }=\widetilde{ \rd_\cA\xi},  \label{CS momentum}
\ees
where $\xi^\mu$ is the scalar field with value in $\cs$.
The physical phase space is therefore given by the quotient of the (infinite dimensional) space of flat connection on $\Sigma$ with the (infinite dimensional)  group of gauge transformations, which action is given by  \eqref{action gauge}. This is the so-called \textit{moduli space of  flat connections}, which happens to be \textit{finite} dimensional \cite{Atiyah:1982fa, Fock:1998nu}.

Since the flatness constraint is actually the same as the moment map, the Poisson structure on moduli space is induced by the quotient, through the Marsden-Weinstein theorem (adapted to deal with infinite dimensional spaces).  

It is actually more convenient to construct the moduli space of flat connections using discretized variables, ie holonomies. Fock and Rosly proposed such construction \cite{Fock:1998nu}. Here we want to propose a new approach which will make the link with gravity more transparent.

\subsection{Chern-Simons discretized 2-form}\label{discrete CS}

The new discretization of the Chern-Simons connection we want to propose follows the same line as the one we used earlier for gravity. In particular, since we do not know how to discretize the density weight 1 variables, we  discretize the variables $\cA$ and expect the discretization to provide us the natural candidates for their discretization.  The discussion in this section is done for any Lie algebra $\cs$. 

\medskip

We consider the triangulation $\Gamma^*$, on which vertices $v$ sit any curvature for the connections $\cA$. Given a face $c^*\in \Gamma^*$, we  consider the  $\CS$ holonomies $H_{ c} $ from the center $c$ in $c^*$ to any other point $x$ in $c^*$. 
We have then   
\be
\cA(x) =(H_c\mone \rd H_c)(x).
\ee
To build the discretized symplectic form, we will use  that 
\be
\delta \cA = \delta( H^{-1}_{c}\rd H_{c}) = H^{-1}_{c}\rd (\delta H_{c} H^{-1}_{c}) H_{c}.
\ee
 The smeared version of 
$\Omega^{\rm CS}$ on the face $c^*$ is then 
\bes\label{PCS}
\Omega^{\rm CS}_{c}&=& \demi \int _{c^*} \,  \la   \rd(\delta H_{c} H^{-1}_{c}) \wedcom  \rd(\delta H_{c} H^{-1}_{c})\ra.\label{bou} 
\ees
Note that unlike the gravity case \eqref{decomp}, there is a symmetry between the two holonomy contributions. 
 
 \medskip
 
Repeating the same steps as in Section \ref{LQG phase space}, introducing $ H_{cc'}=H_{c}(x) H_{c'}(x)\mone$ for $ x\in c\cap c'$, and $ H_{cv}:= H_c(v)$ and $\tilde H^c_{vv'}= H_{cv}\mone H_{cv'} $ we obtain the discretized  symplectic 2-form for the link $\ell=[cc']$. 
\be\label{main result 2}
 \Omega_{cc'}^{\rm CS} =  \demi  \la    H_{cv} \left(\delta \tilde H^c_{vv'} (\tilde H^c_{vv'})^{-1} \right) H_{cv}^{-1}\,  \left(\delta \mH_{cc'} \mH_{cc'}\mone\right) \ra.
 \ee
The holonomies $H_{cv}$ are the connectors mapping the fields from $v$ to $c$. One can see the holonomies discretized on $\tell$ as the natural discretization of the density weight 1 variables $\tcA$. The Levi-Civita tensor accounts for the discretization on the dual of $\ell$.
\be
\tilde  H^c_{\tell} = P\exp\left(\int_\tell  \cA \right) \equiv P\exp\left(\int_\ell \tilde \cA \right).
\ee

\medskip

For a given link $\ell=[cc']$ and its dual $\tell=[vv']$, we build the phase space out of two copies of the  group $\cG$ as a  manifold equipped with $\Omega_{cc'}^{\rm CS}$ as  2-form. To have a phase space, we need this 2-form to be  closed, ie $\delta \Omega_{cc'}^{\rm CS}=0$, which is clearly not the case, due to the presence of the connectors $H_{cv}$. In Section \ref{relations symp}, we are going to see that if $\cG$ is the Euclidian/Poincar\'e group, then $\Omega_{cc'}^{\rm CS}$ is indeed symplectic \textit{up to boundary terms}, so that $\Omega^{discretized}_{\rm CS}=\sum_{\ell}\Omega_{\ell}$ is symplectic.  We leave the study of other groups such as $\SL(2,\C)$, relevant for gravity with a non-zero-cosmological constant, for later studies. 

\medskip
We note that as in the LQG or dual LQG case, all the variables are not independent. Due to the integration around $c^*$, the Chern-Simons holonomy around the face $c^*$ is just the identity.  
\be\label{CS flatness}
\cH_c= \tilde H^c_{v_1v_2}\tilde H^c_{v_2v_3} \tilde H^c_{v_3v_1}=1.
\ee
This is just the statement that there is no torsion or curvature on the face $c^*$, or said otherwise, that the Chern-Simons connection is flat on this face.
\medskip

Given a graph $\Gamma$, with each edge we associate two copies of $\cG$, $\cP_\ell=(\cG_\ell,\cG_\tell)$. When $\Omega_{\ell}^{\rm CS}$ is symplectic (up to boundary terms), we can then define the \textit{kinematical} phase space   of the discretized  Chern-Simons connections on $\Gamma$ by
\be
\cP_{\rm   CS}^{\rm kin}\equiv \times_{\ell\in \Gamma} \cP_\ell/\!/ (\times_{c\in\Gamma} \cH_c), \textrm{ with } \Omega_{\rm CS}^{discrete}=\sum_\ell \Omega_{\ell}^{\rm CS}.
\ee

The physical phase space, ie the moduli space, is obtained by performing the further symplectic reduction with respect to the constraint $\cH_v$ which encodes the flatness constraint (where there are no particles) around the vertices $v$ of $\Gamma^*$. We leave this for further studies.

\subsection{Recovering the Goldman bracket}
\begin{figure}[h]
\centering
\includegraphics[scale=.6]{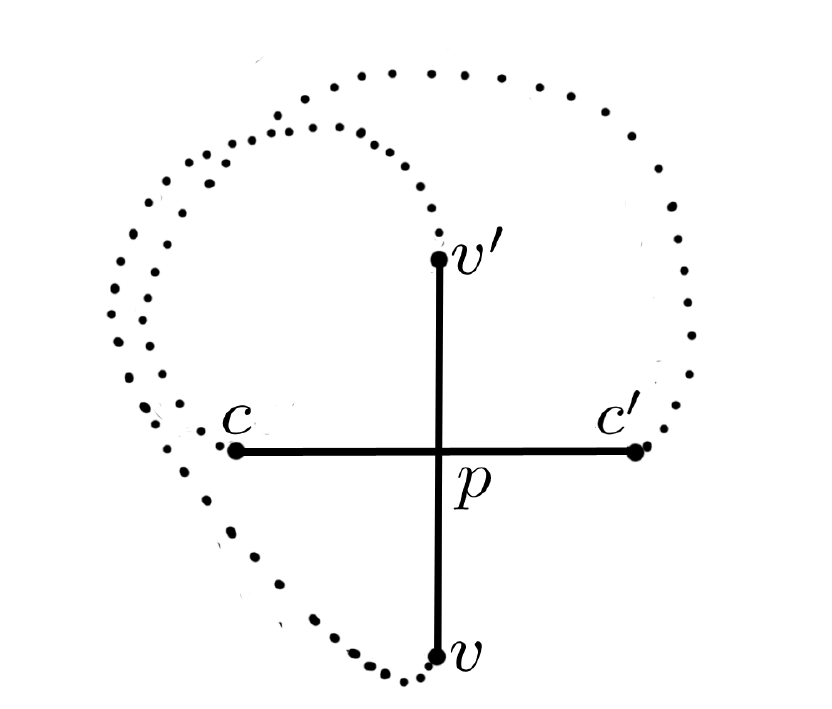}
\caption{The holonomies $H$, $\tilde H$ intersect at $p$.}
\label{goldman}
\end{figure}

We consider the $\CS$ holonomies $H\equiv H_{cc'}H_{c'c}$, $\tH\equiv\tilde H_{v'v} \tilde H_{vv'}$ intersecting at $p$, as sketched in Fig. \ref{goldman}. 

The Goldman bracket is a Poisson bracket between closed holonomies which value is given in terms of the Casimir of $\cs$ \cite{goldman}. 
\be\label{gold-1}
\poi{H^{(1)}, \tilde H^{(2)}}=\tilde{H}_{v'p}^{(2)}H_{cp}^{(1)}\,\cC\, \tilde{H}_{pv'}^{(2)} H_{pc}^{(1)},
\ee
where $\cC=2T^\mu\ot T_\mu$ is the Casimir of $\cs$.

Explicitly, this Poisson bracket can be calculated from the Poisson brackets \eqref{CS poisson} by considering the contributions of  graph intersecting at $p$.
\be\label{gold0}
\poi{H^{(1)}, \tH^{(2)}}=\tH_{v'v}^{(2)} \poi{ H_{cc'}^{(1)},  \tH_{vv'}^{(2)}}H_{c'c}^{(1)},
\ee
and the use of \eqref{CS poisson} gives 
\bes\label{gold1}
\poi{ H_{cc'}^{(1)},  \tilde H_{vv'}^{(2)}}&=&  H_{cp}^{(1)} \tilde H_{vp}^{(2)}\,\cC\, H_{pc'}^{(1)} \tilde H_{pv'}^{(2)},
\ees  
to obtain \eqref{gold-1}. 

The discretized symplectic 2-form we have obtained in \eqref{main result 2} sits at $c$ so it is actually convenient to transport the bracket \eqref{gold1} from $p$ to $c$ to be able to compare them. This is done by  taking advantage of the invariance of the Casimir under the adjoint action. 
\be
\cC=H_{pc}\ot H_{pc}\, \cC \,(H_{pc}\ot H_{pc})\mone.
\ee 
These connectors $H_{pc}$ transport then the right-hand side of \eqref{gold1} to $c$.
\bes
\poi{ H_{cc'}^{(1)},  H_{vv'}^{(2)}}&=&  H_{cp}^{(1)}H_{vp}^{(2)}\,H_{pc}^{(1)} \, H_{pc}^{(2)}\, \cC (H_{cp}^{(1)}\, H_{cp}^{(2)}\, H_{pc'}^{(1)}H_{pv'}^{(2)})\nn\\
&=&   H_{vc}^{(2)}\,  \cC \, H_{cc'}^{(1)}\, H_{cv'}^{(2)} \label{gold2}
\ees 
\smallskip
Let us now reconsider the discretized Chern-Simons symplectic 2-form we have obtained.
\be
 \Omega_{cc'}^{\rm CS} =  \demi \la    H_{cv} \left(\delta \tilde H^c_{vv'} (\tilde H^c_{vv'})^{-1}\right) H_{cv}^{-1}\,  \left(\delta \mH_{cc'} \mH_{cc'}\mone\right) \ra, 
 \ee
 which can be rewritten as 
 \be\label{main result 222}
 \Omega_{cc'}^{\rm CS} =  \demi \la    (H_{cv} \act \,\tilde \theta^R)\,\wedge\,  \theta^R \ra.
 \ee
 $\tilde \theta^R$  and  $\theta^R$ are right invariant 1-form on respectively $\cG_{\tell}$ and $\cG_\ell$. Their dual are the respective right invariant vector fields $\tilde \chi^R$, and $\chi^R$. We have indeed with $H\in \CS$,
  \be
\chi\cdot f(H)= \frac{d}{dt}(f(e^{t\cdot T }H))\left|_{t=0}\right., \quad \chi\cdot H=T^\mu H, \quad 
 \la \theta_\mu\,,\, \chi^\nu\ra=\delta_\mu^\nu.
 \ee
 The dual of   $ H \act \,\theta^R$ is $ H\mone \act \chi^R$. The 2-form $ \Omega_{cc'}^{\rm CS}$ is therefore readily invertible and the associated Poisson bracket (which satisfy the Jacobi identity if and only if  $\Omega_{cc'}^{\rm CS}$ is closed) is given by 
\bes \label{gold3}
\poi{ H_{cc'}^{(1)},  \tilde H_{vv'}^{(2)}}&=& 2\, T^\mu \, H_{cc'}\, \ot \, H_{cv}\mone \,T_\mu \, H_{cv}\, \tilde H_{vv'}\nn\\
&=&H_{vc}^{(2)}\,\cC\, H_{cc'}^{(1)}H_{cv'}^{(2)},
\ees
which is exactly the bracket calculated in \eqref{gold2}. Note that the Poisson bracket we have obtained in \eqref{gold3} does not obviously satisfy the Jacobi identity, due again to the presence of the connectors $H_{cv}$.

Our discretized symplectic 2-form allows therefore to recover the Goldman bracket. Unlike the standard calculation using \eqref{CS poisson}, our Poisson bracket between holonomies is defined for any type of holonomies\footnote{The Poisson bracket of  holonomies sharing some initial/final point is not defined if using  \eqref{CS poisson}. Our discretization scheme regularized this issue.}.

In our formalism, we have recovered a Poisson bracket given it terms of the  Casimir parallel transported which can be seen as a \textit{symmetric} $r$-matrix \cite{semenov}. Hence we have recovered such $r$-matrix by a discretization of the continuum theory. This is to be compared with the   Fock-Rosly formalism \cite{Fock:1998nu} where such structure is put by hand. Furthermore in their framework, the  $r$-matrix contains a non-trivial \textit{anti-symmetric} component which is essential for their construction.

Finally, we emphasize again that our derivation can be extended directly for any $\cs$.

\section{Chern-Simons theory and gravity }\label{sec:CS-grav}
In Section \ref{lqg and dual}, we constructed two kinematical  phase spaces.
The first one is parametrized  by
$(h_{\ell}, \tilde{X}_{\tell}) \in T^*\textrm{G} $ and is associated with each {\it link} $\ell$ of $\Gamma$. The variables are  subject to the Gauss constraints at each node $\cJ_c=0$ where $\cJ_c \in \tilde{\textrm{G}}\sim \mg^*$ 
is in the {\it translational} component of $T^*\textrm{G}$ (ie the momentum component). Some gauge invariant observables, the Wilson loops, are then generated by the {\it curvature} along $\Gamma$: $\prod_{\ell\in \Gamma} h_\ell \in G$.

The second  one  is parametrized by  $(\tilde{h}_{\tell}, {X}_{\ell}) \in T^*\textrm{G} $ associated with each {\it edge} $\tilde{\ell}$ of $\Gamma^*$
and subject to the dual Gauss constraint (ie a flatness constraint) at each vertex $\cG_c=1$ where $\cG_c \in {\textrm{G}}$ 
is in the {\it group} component of $T^*\textrm{G}$. Some gauge invariant observables are then generated by the {\it torsion} along $\Gamma$: $\cJ_v \in \tilde {\textrm{G}}$. This apparent duality can be made more precise, through the notion of \textit{symplectic dual pairs}.

\begin{definition}
A symplectic dual pair between two Poisson manifolds $(X_1, \poi{,}_1)$ and  $(X_2, \poi{,}_2)$ is a correspondance symplectic manifold $(Y, \poi{,}_Y)$ such that, given the Poisson maps $\iota_1$ and $\iota_2$
\be
\begin{array}{ccc}
&Y& \\
\iota_1\swarrow&&\searrow\iota_2\\
X_1&&X_2
\end{array}
\ee
we have for any $f\in\cC(X_1)$ and $h\in \cC(X_2)$,  
\be
\poi{\iota_1^*f,\iota_2^*h}_Y=0.
\ee
\end{definition}
Such precise notion of duality will be useful when considering the quantum versions of our different formulations. The notion of duality then translates into the notion of Morita equivalence, which allows to relate the different representations obtained upon quantization \cite{meyer}.

\medskip

In our context the correspondance symplectic manifold is the phase space generated by the discretized Poincar\'e connections.
To show the symplectic duality,  we are going to show that the symplectic structure on the discretized Poincar\'e connections phase space can be expressed in terms of the each of the symplectic structures of the two LQG phase spaces. 
\be
\Omega^{\rm CS}_{discrete}=\demi\left(\Omega^{\rm LQG}+ \Omega^{\rm LQG^*}\right).
\ee
This is the discrete analogue of the symmetric choice in  \eqref{liouville grav-cs}.

\subsection{ Gravity and Chern-Simons theory}

The gravity action can be related to the Chern-Simons action by a symplectic transformation. 
 Since we are interested in gravity with a zero cosmological constant, we shall consider a Chern-Simons theory built on the Euclidian group or the Poincar\'e group.
 The Poincar\'e (or Euclidian) Lie algebra $\cs$ generated by $T_\mu=(J_A,P_B)$, $A,B=1,2,3$, with brackets 
\beq\label{algebra}
[J_A,J_B]= \epsilon_{ABC}J^C, \quad [J_A,P_B]=\epsilon_{ABC}P^C, \quad [P_A,P_B]= 0, \textrm{ and } J_A^\dagger = -J_A, \, P_A^\dagger = P_A.
\eeq
The indices are raised with the  metric $\eta_{AB}$, the Minkowski or Eucldian metric, according to the choice of spacetime signature. 
A convenient parametrization of $\cs$ is given by setting  \cite{bernd1}
\be\label{theta J}
P_A=\theta J_A, \textrm{ with } \theta^2= 0.
\ee
$\theta$ can be seen as a Grassmanian number, which plays a role similar to the imaginary number $i$. Following \cite{bernd1}, given two real numbers $a,b$, we have 
\be
\overline{(a+\theta b)}= a-\theta b.
\ee
The pairing between the generators  is (see \cite{bernd2} for a discussion on the most general pairing one can consider) is the Killing form of $\cs$.  
\be\la P_A,J_B\ra =\la J_A,P_B\ra = \eta_{AB} =  
-2\int d\theta \,  \tr (J_A P_B) , 
\ee
where $\eta_{AB}$ is the Minkowski (resp. Euclidian) metric if we deal with a Lorentzian (resp. Euclidian) spacetime.  

\medskip

To obtain the gravity variables from the Chern-Simons ones,  we introduce  a pair of connections $\cA_\pm$
\be
\cA_{\pm}\equiv  J_I \omega^{I} \pm P_I e^{I},
\quad
\tilde \cA_{\pm}^a\equiv  \tilde\epsilon^{ba}(J_I  \omega^{I}_b \pm P_I e_b^{I})=J_I \tilde \omega^{Ia} \pm P_I \bfe^{Ia}.
\ee
Note that these connections are not independent since we have that $\cA_+^\dagger= -\cA_-$ and similarly for $\tilde \cA_\pm$. 

We are now interested  in recovering the symplectic 2-forms related to gravity. First we note that 
\be\label{grav-cs-symp}
\Omega^{\rm CS}_\pm=\demi \la \delta\cA_\pm\,\wedge\,\delta \cA_\pm\ra = \pm \demi \left(\la  \delta e \,\wedge\, \delta \omega\ra+ \la \delta \omega\,\wedge\,\delta e\ra\right)= \pm\la  \delta e \,\wedge\, \delta \omega\ra,
\ee
so that the symmetric expression in terms of the gravity variables comes indeed from the Chern-Simons symplectic form as alluded in \eqref{liouville grav-cs}. Conversely we have also 
\bes
\Omega^{\rm LQG}_{grav}&=&  \la \delta e \wedge \delta \omega \ra = \frac1{4}  \la \delta(\cA_{+}- \cA_{-}) \wedge  \delta ( \cA_{+}+\cA_{-}) \ra \nn\\
&=&   \demi (\Omega^{\rm CS}_+- \Omega^{\rm CS}_-).
\ees

\medskip

Finally, we can relate the gravity action to the Chern-simons action.  Using the Killing form based on the Casimir $J^I\ot P_I+P^I\ot J_I $ for the Chern-Simons actions, we have
\be
\cS_{\rm grav}=\demi (\cS_{\rm CS}(\cA_+)- \cS_{\rm CS}(\cA_-)).
\ee

\subsection{Relating  both Chern-Simons and gravity discretized  variables}
As mentioned earlier, we are  interested in the specific case where $\cs=\iso(3)$ or $\iso(2,1)$, that is the gravity case with $\Lambda=0$. 
First we split the Poincar\'e holonomy $H_c=t_c h_c$ into the translational  and rotational parts, respectively $t_c$ and $h_c$. The 
Chern-Simons connection can also be expressed in terms of the gravity variables.
\be
 H_c\mone \rd H_c= h_c\mone (t_c\mone \rd t_c) h_c + h_c\mone \rd h_c = h_c\mone (\ove^KP_K) h_c + \ovo^KJ_K  = \cA = e^K P_K+ \omega ^K J_K .
 \ee
 By construction $t_c$ is an element of $\R^3$ so $\ove$ is a $\R^3$-connection. Furthermore there is only an action of $\su(2)$ on $\R^3$ and no back-action. Therefore, we can identify the connection components  term by term.
\be\label{identification}
\ovo=\omega, \quad  h_c\mone \,\ove\, h_c = e.
\ee
Bearing in mind the discretization of gravity of section \ref{lqg}, it is natural to identify the discretized Chern-Simons components ($h_c, t_c$) with the discretized gravity ones $(g_c,\bfx_c)$ in \eqref{def e}.
\bes
\ovo=\omega &\Rightarrow& h_c \leftrightarrow g_c, \nn\\
 \ees
 Taking advantage of the Grassmannian parameter $\theta$ which gets the exponential  linearized (since $\theta^2=0$), we have  
\beq\label{Yc}
H_{c}= t_{c} g_{c} = e^{ \theta \int_{c}^x d{\bfx}_c } g_{c} = (1+\theta \bfx_{c})g_{c}, \textrm{ with } \bfx_{c}=  \int_{c}^x d{\bfx}_c.
\eeq
We can recover the different fluxes we have introduced when considering the gravity discretization.
\bes
t_{cv}\mone t_{cv'}&=& 1+\theta (\int_{c}^v d{\bfx}_c- \int_{c}^{v'} d{\bfx}_c) = 1+\theta \tilde X_{vv'}\\
t_{c} (h_{cc'} t_{c'}\mone h_{cc'}\mone) &=& 1+\theta (\int_{c}^x d{\bfx}_c- h_{cc'}(\int_{c'}^{x} d{\bfx}_c)h_{cc'}\mone) = 1+\theta (\bfx_c- h_{cc'}\,\bfx_{c'} \,h_{cc'}\mone)=1-\theta  x_{cc'},\nn
\ees
where we used first \eqref{flux1} and then \eqref{xframe}.
 
We can then go further and relate the holonomies entering into the Chern-Simons symplectic form to the (dual) LQG ones. Bearing in mind \eqref{hcvv}, \eqref{Yc}, we have 
\bes
\tilde H_{vv'}&=&H_{cv}\mone H_{cv'} 
=g\mone_{cv}(t_{cv}\mone t_{cv'})g_{cv} g_{cv}\mone g_{cv'}=g\mone_{cv}(t_{vv'})g_{cv}\tilde{h}_{vv'}= g\mone_{cv}(1+ \theta\tilde X^c_{vv'})g_{cv}\tilde{h}_{vv'} \nn\\
&\equiv&  \tilde L _{vv'}\tilde{h}_{vv'}
\ees
In a similar manner, bearing in mind \eqref{gframe}, \eqref{Yc} and \eqref{flux2}
\bes
H_{cc'}&=&H_{c} H_{c'}\mone
=t_{c} g_{c}g_{c'} \mone t_{c'}\mone 
=(t_{c}\,( h_{cc'} \,t_{c'}\mone \, h_{cc'}\mone)) \, h_{cc'} = (1+\theta x_{cc'})  h_{cc'}  \nn\\
&\equiv&  L _{cc'}{h}_{cc'}. 
\ees

Having identified the relationship between the Chern-Simons and gravity discretized variables, it is interesting to see how the Chern-Simons kinematical  flatness constraint \eqref{CS flatness}  from the gravity perspective. A simple calculation shows that it is equivalent to the "kinematical" constraints we have obtained for the LQG and dual LQG variables.  
\be
\tilde H^c_{v_1v_2}\tilde H^c_{v_2v_3}\tilde H^c_{v_3v_1}=1 \Leftrightarrow \tX_{v_1v_2}+  \tX_{v_2v_3}+\tX_{v_3v_1}=0, \quad \tilde h_{v_1v_2}\tilde h_{v_2v_3}\tilde h_{v_3v_1}=1. 
\ee

\subsection{Relating the discretized symplectic 2-forms}\label{relations symp}
We would like to recover the discretized analogue of \eqref{liouville grav-cs}. 
To this aim, it will be useful to redo the full analysis of Section \ref{discrete CS}, while keeping track of the link between the Chern-Simons and gravity  discrete variables.  

\medskip

We start with the Chern-Simons discretized symplectic form on a face $c^*$.
\bes
\Omega^{CS}_{c^*}&=&  \demi\int _{c^*} \, \la d(\delta H_c H_c \mone) \, \wedcom\, d(\delta H_c H_c \mone)\ra ,
\ees
and perform part of the integration  to consider only the integration on the boundary $\partial c^*$ of the face.
\bes\label{55}
\Omega^{CS}_{ c^*}
& =& \demi  \int _{\partial c^*} \,  \la  (\delta H_c H_c \mone) \,\, d (\delta H_c H_c \mone)\ra. 
\ees
We recall that the gravity symplectic form is related to the Chern-Simons symplectic form as given in \eqref{grav-cs-symp}. So we can check now how the discretization affects this, by recovering \eqref{decomp} or \eqref{decomp 2} from \eqref{55}.  We have that  
\be\label{delta G}
\delta H_c H_c\mone = \delta t_{c}t_{c}\mone + t_{c}\, (\delta g_c g_c\mone) \, t_{c}\mone= \delta g_c g_c\mone  +  \theta \left(\delta \bfx_{c} +  [\bfx_{c},\delta g_c g_c\mone ]\right).
\ee
 Plugging this back into \eqref{55}, and restricting $\partial c^*$ to the edge $\tell$, we have
\bes
\Omega^{\rm CS} _{c}&=& \demi  \int _{\tell} \,  \la  (\delta H_c H_c \mone) \,\, d (\delta H_c H_c \mone) \ra\nn\\
&=&
- \int_\tell \la \rd(\delta g_c g_c\mone) \,\, \left(\delta \bfx_{c} +   [\bfx_{c},\delta g_cg_c\mone ]\right) \ra   +  \demi\left[\la (\delta g_cg_c\mone) \,\, \left(\delta \bfx_{c} +   [\bfx_{c},\delta g_cg_c\mone ]\right) \ra \right]_v^{v'} \nn \\
 &=& - \int_\tell \delta\la \rd (\delta g_cg_c\mone) \,\, \bfx_{c}\ra 
 +  \demi\left[\la (\delta g_cg_c\mone) \,\, \left(\delta \bfx_{c} +   [\bfx_{c},\delta g_cg_c\mone ]\right) \ra \right]_v^{v'} \label{CS-lqg 0}\\
&=& \Omega_c + \textrm{ boundary terms.} \nn
\ees
We have recognized in the first term the analogue of  the discretized symplectic form \eqref{decomp 2}, now restricted to the edge $\tell$, whereas the second contribution is a boundary term. Summing up over all of $\partial c^*$, this boundary term  gives 0. 

Since we have  managed to recover $\Omega_c$, we can recover either $\Omega^{\rm LQG}_{cc'}$ or $\Omega^{\rm LQG^*}_{cc'}$ when considering the contributions from the faces $c^*$ and $c'^*$, according to where we put the integration, either on $\delta g_c g_c\mone$ or $\bfx_c$. Hence we have that for the pair of faces $c^*$, $c'^*$
\be
\Omega^{\rm CS}_{cc'}= \demi \left( \Omega^{\rm LQG}_{cc'} + \Omega^{\rm LQG^*}_{cc'}\right) + \textrm{ boundary terms.}
\ee
When considering the full graph $\Gamma$, we have then 
\be
\Omega^{\rm CS}_{discrete}= \demi \left( \Omega^{\rm LQG} + \Omega^{\rm LQG^*}\right).\ee

\section{Outlook}
3d gravity is a nice laboratory to explore some aspects of quantum gravity that could be relevant for 4d gravity. Its connection with Chern-Simons theory also offers rich links with other theories such as  the Wess-Zumino model or conformal theory \cite{carlip}. 
  
 \medskip 
  
In this paper we had a fresh look at the discretization of the 3d gravity with zero cosmological constant, as done in the standard LQG framework. Keeping track of how the continuum variables are related to the discrete variables  allowed to uncover a new discretization, based  on the metric variable (triad). This new framework can be seen as a different choice of polarization, hence in a sense as a dual view of the standard LQG approach.  Although the quantization of the model is still to be done, it could be related to the recent work of Dittrich and Geiller \cite{Dittrich:2014wda, Dittrich:2016typ}. 
  
 \medskip 
  
This fresh look highlighted the fact that non-trivial symmetries are at play, even in the $\Lambda=0$ case. These non-trivial Poisson-Lie group symmetries will lead to quantum group symmetries upon quantization of the theory. Such quantum group symmetries were already identified from different quantized approaches \cite{Freidel:2004nb, Noui:2006ku, Meusburger:2008bs}, but this time we identified them at the classical level.
  
 \medskip 
  
Since the phase space $T^*\SU(2)$ is at the root of LQG both in 3d and 4d,  it is likely that our construction, based on the \textit{metric} formalism, should be generalizable to the 4d picture. Note however that the 3d case is easier than the 4d case, since we essentially deal with graphs ($\Gamma$ and $\Gamma^*$) dual to each other. In 4d, we would have to deal with graphs dual to 2-complexes, which could make the discretization of the holonomy harder. We leave this interesting issue for later.
  
 \medskip 
  
We have used our discretization approach to Chern-Simons theory as well. This provided a new scheme which allowed to recover the Goldman bracket, while keeping a clear link with the continuum variables, unlike the Fock-Rosly formalism. Since we used the same discretizing scheme as for the gravity one, the link between the gravity and the Chern-Simons variables is very clear. Note that such link  between the Fock-Rosly formalism and gravity was discussed in \cite{Meusburger:2015lax, cat-kitaev}. Our new approach to Chern-Simons should be studied further. For example, one should check whether we recover the right  spaces (Heisenberg or Drinfeld doubles) \cite{Alekseev:1993rj, Alekseev:1994pa} when dealing with a punctured space (i.e. particles) or if handles are present. Due to the transparent link with gravity, it would also be  interesting to explore how when introducing boundaries, we can connect the Wess-Zumino model or the conformal theory field theory to Loop (Quantum) Gravity.
  
 \medskip 
  
Finally, we hope also that our approach should shed some light on why we deal with quantum group structures when the cosmological constant is not zero. As we  mentioned already, quantum group structures already appear  when $\Lambda=0$, but from the LQG perspective, there have been many discussions on why we have to deal with a quantum group such as $\SU_q(2)$ (in the Euclidian case) when $\Lambda<0$ \cite{Noui:2011aa, Noui1, Pranzetti1}. Our present work showed that the choice of polarization matters and it is likely that such quantum group will appear when a different polarization than the usual one is chosen. This is work in progress.

\section*{Acknowledgements}
FG would like to thank P. Tiede for discussions. 
This research was supported in part by Perimeter Institute for Theoretical Physics. Research at Perimeter Institute is supported by the Government of Canada through the Department of Innovation, Science and Economic Development Canada and by the Province of Ontario through the Ministry of Research, Innovation and Science.

\appendix
\section{Momentum maps and  symmetries}\label{symplectic}
In this appendix, we review the key-notions of symplectic geometry that are relevant for our results.
There are two different ways to discuss of the notion of symmetries when dealing with symplectic (or more generally Poisson) geometry. In this appendix, we would like to review the different approaches. 

The first one, more mathematically inclined, consists in defining the action $\act$ of a group $G$ on the symplectic manifold $M$. To this aim, it is necessary to equip $G$ with a Poisson structure (which cannot be symplectic by construction) compatible with the group product (so that the action is transitive). Hence $G$ must be a \textit{Poisson-Lie group} \cite{semenov, lu}. We demand then that 
\be \label{math action}
g\act \poi{f_1,f_2}_M=\poi{g\act f_1,g\act f_2}_{G\times M}= \poi{g\act f_1,g\act f_2}_{G}+ \poi{g\act f_1,g\act f_2}_{M}, \quad f_1,f_2\in\cC(M). 
\ee
In physical language, this is stating that a symmetry action on the Poisson bracket of two observables must be the same as the Poisson bracket of the symmetry transformed observables. 

If the Poisson bracket on $G$ is trivial, i.e. 0, as often in standard physics, the only relevant contribution is coming from $M$. However, we can put non-trivial Poisson bracket structures on $G$ and get new realizations of symmetries which were not accessible if setting the Poisson bracket on $G$ to be zero. 

For example, consider the phase space $M=\C^2\ni z_i, i=1,2$, equipped with the following symplectic structure \cite{Shabanov},
\bes
&&\{z_1, z_2\}_M = \frac{i }{\beta}z_1 z_2, \label{first} \quad 
 \{ z_1, \bar{z}_2 \}_M = \frac{i }{\beta}z_1 \bar{z}_2, \label{second} \\
 &&\{z_1, \bar{z}_1 \}_M = -i\left(1-\frac{2}{\beta} (z_1 \bar{z}_1) \right), \label{third} \quad
 \{z_2, \bar{z}_2 \}_M = -i\left(1-\frac{2}{\beta} (z_1 \bar{z}_1 + z_2 \bar{z}_2 ) \right) \label{fourth}
\ees
where $\beta$ is a deformation parameter and all other brackets are zero. 
These Poisson brackets are not covariant under the transformations of $\SU(2)\ni g=\left(\begin{array}{cc}\alpha&-\ov \gamma\\ \ov \gamma&\ov \alpha\end{array}\right)$ with $\det g =1$, unless there is a non-trivial Poisson structure on $\SU(2)$.
\bes
&&\{\alpha, \bar{\alpha}\} = -\frac{2i}{\beta}\gamma\bar{\gamma}, \quad \{\alpha, \gamma \} = \frac{i}{\beta}\alpha\gamma, \quad \{ \alpha, \bar{\gamma} \} = \frac{i}{\beta}\alpha \bar{\gamma},  \quad \{\gamma, \bar{\gamma} \} = 0 . 
\ees
  As such symmetries with non-trivial Poisson bracket can be seen as "hidden symmetries".

\medskip

The second one,  more commonly found in physics,   consists in representing the infinitesimal symmetry transformations using a function on $M$ (with value in a space we are going to determine shortly) and the Poisson bracket on $M$. This function will be called a \textit{momentum map} since for example it is well know that when dealing with $M$ being a cotangent bundle, the momentum coordinates  generate the infinitesimal transformations on configuration space.  

However, these are not the only symmetry transformations one might be interested in. For example, we could be interested in the infinitesimal transformations generated by the configuration space (provided it has a group structure) on momentum space, or it might happen that the phase space $M$ is not of the cotangent type, so that we need to have in hand a formalism that accounts for these general situations. 

\smallskip

Let us note $\mg$ and $\mg^*$, the Lie algebra of $G$ and its dual, the Lie algebra of the group $G^*$. We have also $e_a, e^b$ their respective basis,  and $\chi$ the vector field on $M$ implementing the infinitesimal transformation spanned by the Lie algebra element $x\in \mg$. 
\be
\chi\cdot f= \frac{d}{dt}(e^{tx}\act f)\left|_{t=0}\right..
\ee 
 We define the momentum map $\cP$ which will implement the infinitesimal transformation using the Poisson bracket on $M$ \cite{Babelon2, Alekseev:yg}. 
\be
\begin{array}{rcl}
\cP: M &\dr& G^* \\
 \xi &\dr &g_*(\xi)= e^{-Q(\xi)} \textrm{ such that }  \chi\cdot f = \la g_*\mone \poi{f,g_*}_M,x\ra, \quad f\in\cC(M)
\end{array}
\ee
where $g_*\mone$ is the inverse of the group element in $G^*$ and $\la,\ra$ is the bilinear form between $\mg$ and $\mg^*$. $Q(\xi)$ is a $\mg^*$-valued function which is called the \textit{charge} generating the Poisson-Lie group action. Note that in this definition, $\chi$ is a right-invariant vector; a similar definition would hold for the left-invariant one. 

We can connect this definition to the more abstract definition of the notion of symmetry action \eqref{math action}.  By considering the infinitesimal version of \eqref{math action}, bearing in mind that the infinitesimal version of the Poisson bracket on $G$ is given by the structure constant $C_a^{bc}$ of $\mg^*$, we get that 
\be
e_a\cdot \poi{f_1,f_2}_M= C_a^{bc}(e_b\cdot f_1)(e_c\cdot f_2)+ \poi{e_a \cdot f_1,f_2}_{M}+ \poi{f_1,e_a\cdot f_2}_M.
\ee
On the other hand, using the definition of the symmetry action involving  the momentum map we recover the same expression. 
\bes
&&\chi\cdot \poi{f_1,f_2} - \poi{\chi \cdot f_1,f_2}- \poi{f_1,\chi\cdot f_2}= \\ && \la g_*\mone\poi{\poi{f_1,f_2},g_*} -\poi{g_*\mone\poi{f_1,g_*},f_2}-\poi{f_1,g_*\mone \poi{f_2,g_*}} , X\ra \nn = \la \left[g_*\mone \poi{f_1,g_*}, g_*\mone\poi{f_2,g_*}\right], X\ra
\ees
where we used the Jacobi identity, and the last term provides the contribution $C_a^{bc}$ accounting for the non-trivial Poisson structure on $G^*$.

\medskip 

Let us illustrate this construction when we deal with $M=T^*G\sim G\, \act\!\!\! < \mg^* $ (or example $G=\SU(2)$), with coordinates $(X,g)$ in the left trivialization. In this case, $G^*\sim \mg^*\sim \R^3$ is an abelian group. We use the Poisson bracket 
\be
\poi{X_i,X_j}=\epsilon^{k}_{ij}X_k, \quad \poi{X_i,g}=J_i g, \quad \poi{g,g}=0.
\ee
We note therefore that $G$ has a trivial Poisson structure whereas $G^*\sim\mg^*$ is equipped with a  non-trivial one, specified by the Lie algebra structure constants $\epsilon^{k}_{ij}$  of $\mg$.

\smallskip

Since we have two groups, $G$ and $G^*$, we can construct two types of momentum maps. The standard one $\cJ$, with value in $G^*\sim \mg^*$ and a may be less usual one $\cG$ with value in $G$. The former one will implement the infinitesimal rotations on configuration space, while the latter one will implement the infinitesimal translations (since $G$ is non-abelian, there will be a difference between left and right) on $G^*$.  

We define then (we recall we use the left trivialization of $T^*G$ and we note respectively $P_i$ and $J_j$ the generators of the Lie algebras $\R^3$ and $\mg$, such that $\la P_i,J_j\ra=\delta_{ij}$), 
\bes
\begin{array}{rcl}
\cJ: T^*G&\dr& \mg^*\sim G^*\sim \R^3 \\
(X,g)&\dr & g_*(X)=e^{-X^iP_i}
\end{array}
\quad 
\begin{array}{rcl}
\cG: T^*G&\dr&  G\\
(X,g)&\dr & g
%=e^{-\beta^iJ_i}
\end{array}
\ees
The infinitesimal transformations they implement are respectively ($\beta^i\in\R^3$)
\bes
&&\chi^i_\cJ \cdot X^j = \la g_*\mone \poi{X^j,g_*},J^i\ra=\epsilon^{ijk}X_k , \quad \chi^i_\cJ \cdot g =\la g_*\mone \poi{g,g_*},J^i\ra =-J^ig \\
&& \beta_i \chi^i_\cG \cdot X^j = \la g\mone \poi{X^j,g},\beta_i P^i\ra= g\act \beta^i 
, \quad \beta_i \chi^i_\cG \cdot g = \la g\mone \poi{g,g},\beta_i P^i\ra=0, \label{translation}
\ees
 where $g\act \beta^i$ denotes the coadjoint action\footnote{Hence if $G=\SU(2)$, $\beta= \beta^i J_i$, then $g\act \beta= g\mone \beta g$.} of $g$ on the vector $\beta^i$.

\medskip

A phase  phase  based on the manifold $M= G^*\bowtie G$ is called the Heisenberg double \cite{lu, Alekseev:1993np, semenov}. The cotangent bundle  $T^*\SU(2)\sim SU(2)\act\!\!\!\!<\R^3 $ is a simple example of such structure.   The general Poisson Lie symmetry of the Heisenberg double is given by the Drinfeld double $D=G^*\bowtie G$, which acts by left or right translation on $M$. As we discussed in our simple example, due to the symmetry between $G$ and $G^*$, we can build momentum maps with value in $G$ or $G^*$. The general momentum map associated to the Drinfeld double has value in $D^*\sim D$.

\medskip
In our construction, we are interested in putting together many copies of the same Heisenberg double $M$, and considering some global symmetry transformations. The momentum map is then extended  using the group product \cite{Alekseev:yg}. First let us recall that the Poisson structure on $\cC(M^{\times n})\sim \cC(M)^{\ot n}$ is given by the sum of the Poisson bracket on each of the individual components $\cC(M)$. If $G$ acts on $M^{\times n}$, then we define the associated momentum map and the  global infinitesimal  (right) action by 
\be
\begin{array}{rcl}
\cP_{tot}: M^{\times n} &\dr& G^* \\
 (\xi_1,..\xi_n) &\dr &g_*(\xi_1)..g_*(\xi_n) =g_{*tot}\textrm{ such that }  \chi_{tot}\cdot f = \la g_{*tot}\mone \poi{f,g_{*tot}}_{M^{\times n}},x\ra,\end{array}
\ee
with $f\in\cC(M^{\times n})$.

\bibliographystyle{unsrt}

\bibliography{biblio}{}

\end{document}